# Boron-graphdiyne: superstretchable semiconductor with low thermal conductivity and ultrahigh capacity for Li, Na and Ca ions storage

Bohayra Mortazavi[*,1], Masoud Shahrokhi[2], Xiaoying Zhuang[3] and Timon Rabczuk[4, #]

[1]Institute of Structural Mechanics, Bauhaus-Universität Weimar, Marienstr. 15,
D-99423 Weimar, Germany.
[2]Institute of Chemical Research of Catalonia, ICIQ, The Barcelona Institute of Science and Technology,
Av. Països Catalans 16, ES-43007 Tarragona, Spain.
[3]Institut für Kontinuumsmechanik, Gottfried Wilhelm Leibniz Universität Hannover,
Appelstrasse 11, 30167 Hannover, Germany.
[4]College of Civil Engineering, Department of Geotechnical Engineering, Tongji University, Shanghai, China.

## Abstract

Most recently, boron-graphdiyne, a π-conjugated two-dimensional (2D) structure made from merely sp carbon skeleton connected with boron atoms was successfully experimentally realized through a bottom-to-up synthetic strategy. Motivated by this exciting experimental advance, we conducted density functional theory (DFT) and classical molecular dynamics simulations to study the mechanical, thermal conductivity and stability, electronic and optical properties of single-layer B-graphdiyne. We particularly analyzed the application of this novel 2D material as an anode for Li, Na, Mg and Ca ions storage. Uniaxial tensile simulation results reveal that B-graphdiyne owing to its porous structure and flexibility can yield superstretchability. The single-layer B-graphdiyne was found to exhibit semiconducting electronic character, with a narrow band-gap of 1.15 eV based on the HSE06 prediction. It was confirmed that the mechanical straining can be employed to further tune the optical absorbance and electronic band-gap of B-graphdiyne. Ab initio molecular dynamics results reveal that B-graphdiyne can withstand at high temperatures, like 2500 K. The thermal conductivity of suspended single-layer B-graphdiyne was predicted to be very low, ~2.5 W/mK at the room temperature. Our first-principles results reveal the outstanding prospect of B-graphdiyne as an anode material with ultrahigh charge capacities of 808 mAh/g, 5174 mAh/g and 3557 mAh/g for Na, Ca and Li ions storage, respectively. The comprehensive insight provided by this investigation highlights the outstanding physics of B-graphdiyne nanomembranes, and suggest them as highly promising candidates for the design of novel stretchable nanoelectronics and energy storage devices.

Corresponding authors: *bohayra.mortazavi@gmail.com

## 1. Introduction

Since the successful isolation of graphene [1,2], an immense amount of research has been conducted in the field of two dimensional (2D) materials. Graphene, the most well-known member of 2D materials shows exceptional physics, including interesting electronic and optical properties along with highest measured mechanical strength [3] and thermal conductivity [4]. These outstanding material properties propose the graphene as a highly attractive candidate for diverse applications, such as; heat management components, mechanically robust and stretchable nanodevices, nanoelectronics and nanooptics. Graphene nevertheless yields zero band-gap semiconducting electronic character in its defect-free form, and thus it is not suitable for the application as a 2D transistor. This electronic nature of graphene however has been acting positively, promoting the design and discovery of novel 2D materials with inherent semiconducting electronic properties

Among various classes of 2D materials with semiconducting electronic character, during the last few years the experimental realization of carbon based 2D semiconductors has attracted remarkable attention. In this regard, covalent networks of carbon and nitrogen atoms arranged in lattices with different atomic compositions have been among the most successful experimental accomplishments. Graphitic carbon nitride g-$C_3N_4$ [5,6] are well-known 2D semiconductors that have been experimentally fabricated, with desirable performances for various applications, like; energy storage and conversion, fuel cells, catalysis, photocatalysis and $CO_2$ capture [5,7–12]. Nitrogenated holey graphene is another attractive member of carbon-nitride 2D materials family, which was successfully synthesized via a simple wet-chemical reaction [13]. Recently, a graphene-like 2D polyaniline crystals with $C_3N$ stoichiometry was experimentally realized via the direct pyrolysis of hexaaminobenzene trihydrochloride single crystals in solid state [14]. Carbon-nitride 2D structures provide very attractive properties, suitable for post-silicon nanoelectronics [13–19].

Graphyne [20] structures are another class of planar full carbon allotropes, including sp and $sp^2$ hybrid bonded atoms arranged in various crystal lattices. Interestingly, in 1987 Baughman *et. al* [20] predicted numerous graphyne structures, some with semiconducting electronic character. Amazingly, three decades after this original theoretical work, two graphyne structures have been recently experimentally synthesized. In 2017 Jia *et al.* [21] reported the fabrication of carbon Ene-yne graphyne from tetraethynylethene by solvent-phase reaction. Shortly after, the synthesize of crystalline graphdiyne nanosheets were reported by Matsuoka *et al.* [22]. The appealing physics of graphyne structures have been an attractive topic for theoretical studies. These full carbon nanomembranes have been theoretically predicted to yield highly promising properties for diverse applications, such as; anode material for metal-ion batteries [23,24], hydrogen storage [25–28], catalysts [29], thermoelectricity [30,31] and nanotransistors [32–35].

Most recently, an exciting experimental advance has just taken place with respect to the synthesis of B-graphdiyne 2D structure through a bottom-to-up synthetic strategy [36]. The fabricated B-graphdiyne structure by Wang *et al.* [36] is likely to that of graphene-like graphyne; however, in B-graphdiyne single boron atoms replace the connecting hexagonal carbon rings in the graphyne lattice. This latest experimental success in the fabrication of B-graphdiyne highlights the importance of theoretical studies in order to provide understanding of its intrinsic material properties. Such that comprehensive analysis of structural, thermal, mechanical, electronic and optical properties of 2D B-graphdiyne structures plays critical roles in the design of advanced nanodevices exploiting the outstanding properties of this novel material. The objective of the present investigation is therefore to efficiently explore the material properties of B-graphdiyne structure through extensive atomistic simulations. To this aim, we conducted extensive first-principles density functional theory (DFT) simulations to investigate mechanical, thermal stability, optical and electronic properties of this newly synthesized 2D structure. The phononic thermal conductivity was also predicted using the classical non-equilibrium molecular dynamics simulations. It is worthy to note that 2D materials and heterostructures owing to their good stability, desirable adsorption energy, remarkable ionic conductivity and high storage capacity have recently attracted remarkable attention for the application as active materials in rechargeable metal ion batteries [37–45]. We therefore also employed the DFT calculations to investigate the possible application of B-graphdiyne as an anode material for Li, Na, Mg and Ca ions storage. This work provides a comprehensive vision concerning the critical properties of a novel class of 2D semiconductors and hopefully the acquired results can guide future theoretical and experimental studies.

## 2. Computational methods

In this work we employed Vienna *Ab initio* Simulation Package (VASP) [46–48] to conduct the density functional theory simulations with the Perdew-Burke-Ernzerhof (PBE) functional [49] for the exchange correlation potential. The interaction between the valence and core electrons was described on the basis of the projected augmented wave (PAW) method [50]. A plane-wave cutoff energy of 500 eV was used for the valence electrons. The VESTA [51] package was utilized for the visualization of atomic structures. In Fig.1, the energy minimized B-graphdiyne lattice with a chemical composition of $C_{12}B_2$ which was realized experimentally by Wang *et al.* [36] is illustrated. As it is clear, B-graphdiyne shows a graphene-like lattice. This way in analogy to graphene and in order to analyze the anisotropicity in transport properties, we examined the thermal and mechanical properties along the armchair and zigzag directions, as depicted in Fig. 1. For the evaluation of optical responses, the *x* and *y* directions (as shown in Fig. 1) were however considered, according to the unit-cell vectors. In all simulations, periodic boundary conditions were applied along all three Cartesian directions with a vacuum thickness of 20 Å, to avoid image-image interactions along the sheet normal direction.

In order to evaluate the mechanical properties, we conducted uniaxial tensile simulations. In this case, only a unit-cell structure was considered. We analysed the anisotropy in the mechanical properties by conducting the uniaxial tensile simulations along the armchair and zigzag directions. To this aim, the periodic simulation box size along the loading direction was increased gradually with a fixed engineering strain. In order to satisfy the uniaxial tensile loading condition, the stress along the transverse directions of loading should remain negligible. Since the atoms are in contact with the vacuum along the sheet's normal direction, the stress in this direction is always negligible. Therefore, to satisfy the uniaxial tensile loading condition the simulation box size along the sheet's width was altered to also reach a negligible stress. In order to simulate the atomic rearrangements during the tensile simulations, the conjugate gradient method was employed for the geometry optimizations, with termination criteria of $10^{-4}$ eV and 0.01 eV/Å for the energy and the forces, respectively, using a 7×7×1 Monkhorst-Pack [52] k-point mesh size. To evaluate the thermal stability, ab initio molecular dynamics (AIMD) simulations were carried out for a rectangular unit-cell, using the Langevin thermostat with a time step of 1 fs and 2×2×1 k-point mesh size.

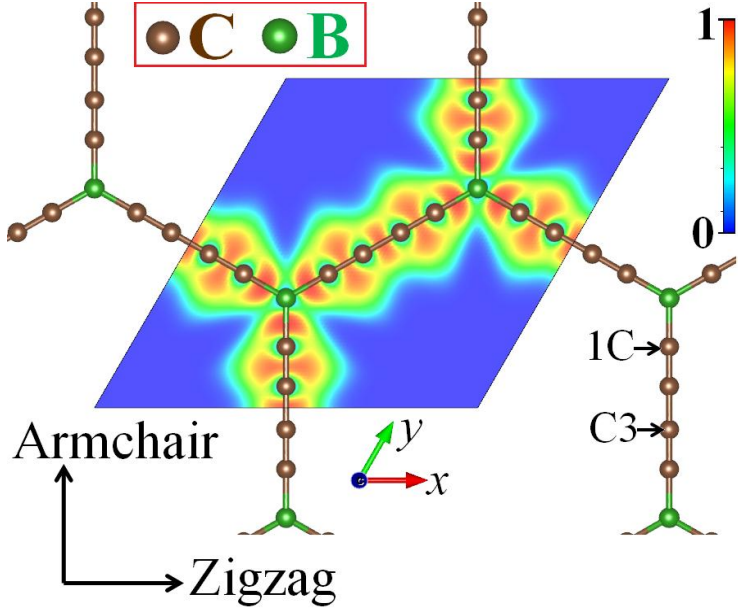

**Fig. 1**, Atomic structure of $C_{12}B_2$ B-graphdiyne. Contour illustrates the electron localization function within the unit-cell. Transport properties were studied along the armchair and zigzag directions as shown.

Since the PBE functional underestimates the band-gap values, we also employed the screened hybrid functional, HSE06 [53] to evaluate the electronic properties of $C_{12}B_2$ monolayer. We used the Gaussian smearing method with smearing width of 0.02 eV and 8×8×1 k-point grid for computing the optical properties. Optical properties, including the imaginary and real parts of dielectric and absorption coefficient were calculated through the random phase approximation (RPA) method [54]. The optical spectra of the single-layer $C_{12}B_2$ was acquired using the PBE plus RPA approach. Optical properties were described by photon frequency dependent dielectric function, $\varepsilon(\omega) = \operatorname{Re}\varepsilon_{\alpha\beta}(\omega) + i\operatorname{Im}\varepsilon_{\alpha\beta}(\omega)$. The imaginary part of the dielectric function for semiconductors could be obtained by taking into account the contribution of interband transition contribution [55,56]:

$$\operatorname{Im}\varepsilon_{\alpha\beta}(\omega) = \frac{4\pi^2 e^2}{\Omega} \lim_{q\to 0} \frac{1}{|q|^2} \sum_{c,v,k} 2w_k \delta(\varepsilon_{ck} - \varepsilon_{vk} - \omega) \times \langle u_{ck+e_\alpha q}|u_{\alpha k}\rangle \langle u_{ck+e_\beta q}|u_{\alpha k}\rangle^* \quad (1)$$

In this equation, $q$ is the Bloch vector of the incident wave and $w_k$ is the **k**-point weight. The band indices $c$ and $\alpha$ are restricted to the conduction and the valence band states, respectively. The vectors $e_\alpha$ are the unit vectors for the three Cartesian directions and $\Omega$ is the volume of the unit-cell. $u_{ck}$ is the cell periodic part of the orbitals at the $k$-point $\mathbf{k}$. The real part $\text{Re}\varepsilon_{\alpha\beta}(\omega)$ can be evaluated from $\text{Im}\varepsilon_{\alpha\beta}(\omega)$ using the Kramers–Kronig transformation [55]:

$$\text{Re}\,\varepsilon_{\alpha\beta}(\omega) = 1 + \frac{2}{\pi} P \int_0^\infty \frac{\omega' \text{Im}\,\varepsilon_{\alpha\beta}(\omega')}{(\omega')^2 - \omega^2 + i\eta} d\omega' \quad (2)$$

where $P$ denotes the principle value and $\eta$ is the complex shift in Kramers-Kronig transformation. The absorption coefficient was calculated using the following relation [57]:

$$a_{\alpha\beta}(\omega) = \frac{2\omega k_{\alpha\beta}(\omega)}{c} \quad (3)$$

where $k_{\alpha\beta}$ is the imaginary part of the complex refractive index and $c$ is the speed of light in vacuum, known as the extinction index. $k_{\alpha\beta}$ was acquired according to:

$$k_{\alpha\beta}(\omega) = \sqrt{\frac{\left|\varepsilon_{\alpha\beta}(\omega) - \text{Re}\,\varepsilon_{\alpha\beta}(\omega)\right|}{2}} \quad (4)$$

The optical spectra of B-graphdiyne nanosheets have been obtained for the in-plane directions in order to assess the anisotropicity of optical properties.

We employed classical non-equilibrium molecular dynamics (NEMD) method to predict the lattice thermal conductivity of B-graphdiyne. The NEMD simulations in this study were conducted using the LAMMPS [58] package. In this work the atomic interactions were introduced by employing the Tersoff potential [59] with the optimized parameters by Lindsay and Broido [60] and Kinarci *et al.* [61] for carbon-carbon and carbon-boron interactions, respectively. Worthy to note that the employed potential functions have been widely used in the classical molecular dynamics modelling of heat transfer in carbon based 2D materials [62–65]. In all NEMD simulations, we applied periodic boundary condition along the planar directions and a small time increment of 0.1 fs was used. NEMD calculations were performed for B-graphdiyne samples with different lengths to probe the length effect on the predicted thermal conductivities. To evaluate the thermal conductivity, we first relax the structures at the room temperature using the Nosé-Hoover thermostat method (NVT). We then fixed atoms at the two ends of the sample and divided the simulation box (excluding the fixed atoms) along the heat transfer direction into 20 slabs. The first two slabs at the two ends were assigned to be cold and hot slabs, respectively. To simulate the heat transfer, we applied a 20 K temperature difference between the hot and cold slabs. During the NEMD simulations, the temperatures in the hot and cold slabs were controlled by the NVT method, while the constant energy, NVE ensemble with velocity-Verlet integration method was used to simulate the atomic motions in the remaining slabs. These applied loading conditions, impose a heat-flux along the heat transfer direction by continuously adding small amounts of energy to the atoms in the hot slabs and simultaneously removing energy from the atoms in the cold slabs. The applied heat-flux, $q_x$ was calculated assuming a thickness of 3.35 Å for single-layer $C_{12}B_2$, the same as that of the graphene. As a result of the imposed heat flux in the system, a steady-state linear temperature profile establishes along the sample. The thermal conductivity, $\kappa$, was then evaluated using the established temperature gradient and the applied heat-flux, according to the one-dimensional form of the Fourier law:

$$\kappa = q_x \frac{dx}{dT} \qquad (5)$$

For the simulations of Li, Na, Mg and Ca adatoms adsorption over the unit-cell structure of single-layer B-graphdiyne, a dispersion scheme of DFT-D2 [66] was employed to improve the binding energy calculations by accounting for the dispersion corrections. In this case to find the minimized structure, the conjugate gradient method was employed with convergence criteria of $10^{-4}$ eV and 0.01 eV/Å for the energy and the forces, respectively, using a 5×5×1 Monkhorst-Pack [52] k-point mesh size. To precisely calculate the final energy values and charge densities, we conducted single point calculations over the energy minimized structures, using the tetrahedron method with Blöchl corrections [67] in which the Brillouin zone was sampled with a 11×11×1 k-point mesh size. The nudged elastic band (NEB) method was employed to simulate the diffusion of a single Li adatom.

## 3. Results and discussions

In the Fig. 1, top view of the energy minimized graphene-like B-graphdiyne monolayer is illustrated. For the energy minimized structure, the hexagonal lattice constant was measured to be 11.847 Å. In this structure, only three different types of atoms exist; B atoms and C atoms connected with either B or C atoms. In accordance with the original experimental work by Wang et al. [36], C atoms connected with B and C atoms are distinguished with 1C and C3 terms, respectively. In the energy minimized structure, the B-1C, 1C-C3 and C3-C3 bond lengths were measured to be 1.512 Å, 1.235 Å and 1.345 Å, respectively. As it is clear, B-C bonds exhibit the maximum length which may imply their lower rigidity as compared with other bonds in this system. The atomic structure of energy minimized single-layer B-graphdiyne is provided in the supporting information of this manuscript. In Fig. 1, we also plotted the electron localization function (ELF) [68], which takes a value between 0 and 1, where ELF=1 corresponds to the perfect localization. As it is clear from the ELF contour calculated for the $C_{12}B_2$ monolayer, the electron localization occurs on the center of the C-B and C-C bonds, which confirms the covalent bonding in this structure.

We investigated the energetic stability of B-graphdiyne monolayer. To this aim, the cohesive energy per atom was calculated as defined by $E_{coh} = -(\sum_i E_i - E_t)/n$, where $E_t$, $E_i$ and $n$ are the total energy of the system, the energy of the i-th isolated atom and the total number of atoms in the unit-cell, respectively. The cohesive energy of single-layer B-graphdiyne was calculated to be -5.85 eV, which is higher than that of the graphene (-9.23 eV [69]). The negative cohesive energy for this monolayer suggests that the free-standing single-layer B-graphdiyne is energetically stable.

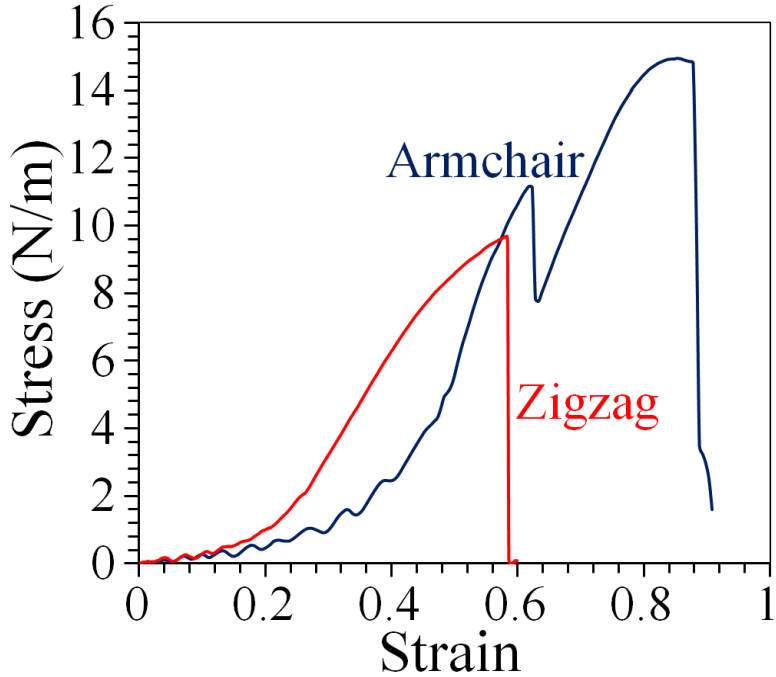

**Fig. 2**, First-principles predictions for uniaxial stress-strain responses of single-layer graphene-like B-graphdiyne along the armchair and zigzag directions.

3.1 Mechanical response

We first analyze the mechanical response of B-graphdiyne by conducting the uniaxial tensile simulations. In Fig. 2, the first-principles results for the uniaxial stress-strain responses of single-layer $C_{12}B_2$ elongated along the armchair and zigzag directions are illustrated. For the most of materials with densely packed structures, the uniaxial stress-strain curves show initial linear responses corresponding to the linear elasticity. Notably, the uniaxial stress-strain responses of B-graphdiyne include no initial linear relations and thus this system does not exhibit linear elasticity. Worthy to note that the linear elastic response in the mechanical properties is usually associated with the bond stretching as a result of applied mechanical strains. As an example for the case of pristine graphene, the stretching of the structure can be achieved only by increasing the carbon atoms bond lengths and such that at initial strain levels the stress values increase linearly [70]. As it can be seen from the results shown in Fig. 2, by increasing the strain levels the strain values also increase but in sinusoidal patterns. This observation can be attributed to the fact that at initial strain levels the stretching of B-graphdiyne does not directly leads to the increase of bonds lengths. This way, in this novel 2D system the deformation is achieved by a combination of structural deflection and bond stretching. The dominance of bond stretching in the deformation process correlates with the increase in the stress values. On the other side the decreases in the stress values reveal that the structural deflection, which is mainly achieved by the contraction of the structure along the perpendicular direction of loading (sheet's width).

Uniaxial stress-strain relations illustrated in Fig. 2 confirm that along the armchair direction the B-graphdiyne structure yields distinctly higher stiffness and stretchability as compared with the zigzag direction. The ultimate tensile strength of single-layer $C_{12}B_2$ along the armchair and zigzag directions were predicted to be 15 N/m and 9.6 N/m, respectively. This can be explained because of the fact that for the uniaxial loading along the armchair direction from every two carbon-carbon chains existing in the system, one is exactly oriented along the loading direction and thus the bonds are more involved in the load transfer and stretching. Along the armchair direction the B-graphdiyne exhibits around 50% higher stretchability as compared with zigzag direction. We remind that for the uniaxial loading along the armchair and zigzag directions, the sinusoidal stress patterns are observable up to the strain levels of ~0.4 and ~0.12, respectively. In these cases, as the loading proceeds the bond elongations are observable by the increase of the stress values, however as considerable contraction along the sheet's width occurs, some parts of the stresses in the bonds relieve, resulting in overall decline in the effective stress values. This way, the higher stretchability along the armchair direction can be explained by the more contraction ability of the structure along the perpendicular direction of loading. According to our first-principles modelling results, the maximum strains that the graphene-like B-graphdiyne can keep its load bearing ability, equivalent with the stretchability, were predicted to be to be ~0.88 and ~0.58 for the uniaxial loading along the armchair and zigzag directions, respectively. Worthy to mention that the strain at the ultimate tensile strength point for pristine graphene and hexagonal boron-nitride were found to be ~0.27 [70] and 0.3 [71], respectively. Interestingly, the stretchability of B-graphdiyne is also considerably higher than other 2D full carbon porous networks [72], carbon Ene-yne [73] and graphyne [74] structures. To the best of our knowledge, the graphene-like B-graphdiyne outperforms other planar 2D materials with respect to the stretchability.

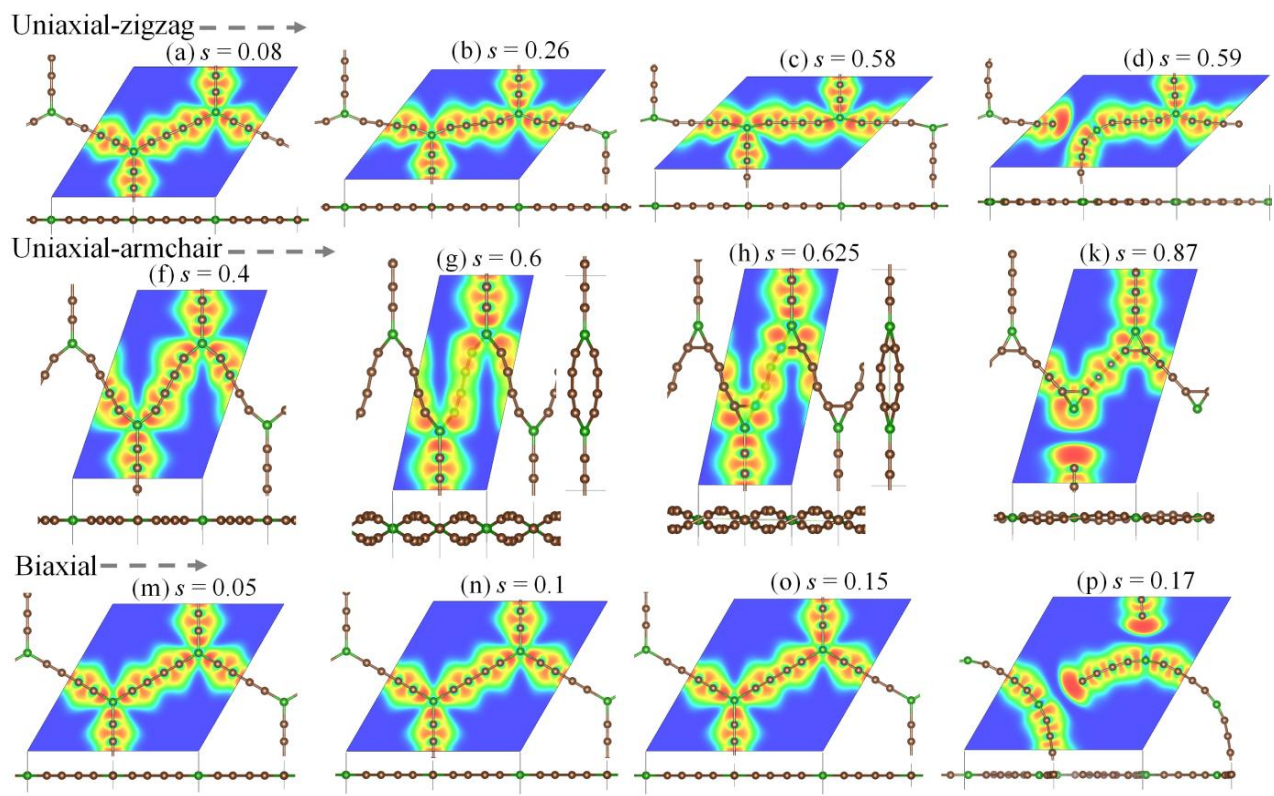

**Fig. 3**, Top and side views of the deformation process of B-graphdiyne monolayer under different loading conditions and at different strain levels ($s$). The length criteria for the bonds illustration was chosen to be 2 Å.

In order to briefly analyze the failure mechanism of $C_{12}B_2$ B-graphdiyne, snapshots of the deformation process under different loading conditions and at different strain levels along with the ELF contours are compared in Fig. 3. In this case, we also conducted the biaxial tensile simulation to compare the results with those of the uniaxial loadings. According to the results shown in Fig. 3, for the all studied samples the bond breakage happen in B-C bonds. The bond breakages can be also understood from the ELF contours, when the electron localization vanishes from the center of original bonds and concentrates around the atoms. For the samples under the uniaxial tensile loading conditions, the bond breakages happen for those oriented along the loading direction, those were involved in the stretching and load bearing. For the uniaxial loading along the zigzag direction, the carbon-carbon chains initially oblique to the loading direction with an angle of 30 degree (Fig. 3a), during the uniaxial loading rotate and finally become in-line with the loading direction (Fig. 3c). In this case, during the entire deformation process the structure retains its original planar structure. For the uniaxial loading the armchair direction, considerable contraction of the structure along the perpendicular direction of the loading is observable. In this case, until the strain level of ~0.4, which is associated with the strain limit for sinusoidal stress pattern, the planarity of the structure is preserved (Fig. 3f). After this point, by increasing the strain level the carbon-carbon chains initially oblique to the loading direction by an angle of 60 degree tend to orient along the loading direction and simultaneously start to deflect in the out-of-plane direction (as shown in Fig. 3g). Nonetheless carbon-carbon chains originally oriented along the loading direction are kept completely in-plane (Fig. 3g side views). As illustrated in Fig. 2, the stress-strain curve for the uniaxial loading along the armchair direction reveals an unusual first yield point at the strain level of ~0.6, in which a conspicuous sudden drop in the stress value is observable. According to our results shown in Fig. 3h, at this initial yield point the carbon-carbon chains get close to each other and suddenly form new C-C bonds, resulting in the relieve of some parts of the stresses. After this point and by further increasing the strain level, the stress values increase sharply again. This increase is due to the fact that the monolayer ability to contract along the transverse direction of loading becomes considerably restricted and thus the subsequent deformation can be only accomplished by the bond stretching. As shown in Fig. 3k, these newly formed C-C bonds construct triangular chains that can stay intact in the structure, even after the occurrence of the final rupture. The analysis of deformation process clearly explain the higher stretchability and tensile strength of graphene-like B-graphdiyne along the armchair direction. In a simpler word, as observable in Fig. 3c, for the uniaxial loading along the zigzag only half of the carbon-carbon chains in the system can contribute in the load bearing and stretching as well, whereas for the uniaxial loading along the armchair, all carbon-carbon chains finally involve in the load bearing (see Fig. 3h). Worthy to note that for the biaxial loading, since the deformation can be achieved only by the bond elongation, the stretchability is highly limited. In this case, as expected and likely to the samples under the uniaxial loadings, the ruptures occur in the B-C bonds (Fig. 3p).

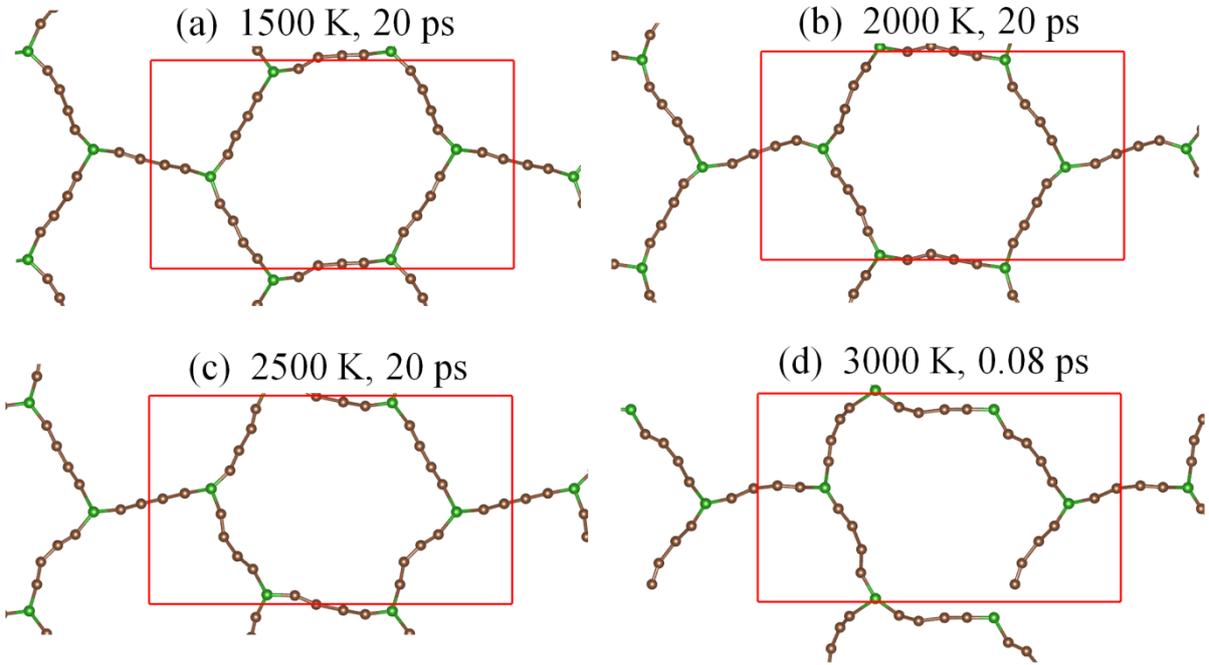

**Fig. 4**, Snapshots of single-layer $C_{12}B_2$ B-graphdiyne at different temperatures of 1500 K, 2000 K, 2500 K and 3000 K. The AIMD results of 20 ps long simulations confirm the outstanding thermal stability of graphene-like B-graphdiyne up to 2500 K. The length criteria for the bonds illustration was chosen to be 2 Å.

### 3.2 Thermal stability

The range of temperatures that a material can endure intact is another important property that plays a critical role for the high temperature applications. We therefore examined the thermal stability of single-layer B-graphdiyne using the AIMD simulations at high temperatures. To this goal, AIMD simulations were conducted at different temperatures of 500 K, 1000 K, 1500 K, 2000 K, 2500 K and 3000 K for up to 20 ps. The snapshots of the single-layer $C_{12}B_2$ after the AIMD simulations are illustrated in Fig. 4. As a remarkable finding, the B-graphdiyne monolayer was found to stay intact at the high temperature of 2500 K (Fig. 4c). This novel 2D system is however rapidly disintegrated at the higher temperature of 3000 K (Fig. 4d). In this temperature, in accordance with failure mechanism under the mechanical loading, the first bond breakage occurs in the B-C bonds (see Fig. 4d). Our AIMD results reveal the outstanding thermal stability of graphene-like B-graphdiyne and thus confirms its suitability for service at high temperatures. Worthy to note that the outstanding thermal and mechanical stability of B-graphdiyne monolayer does not guarantee its dynamical stability, such that the analysis of phonon spectrum of this novel 2D system can be an important topic for the future studies.

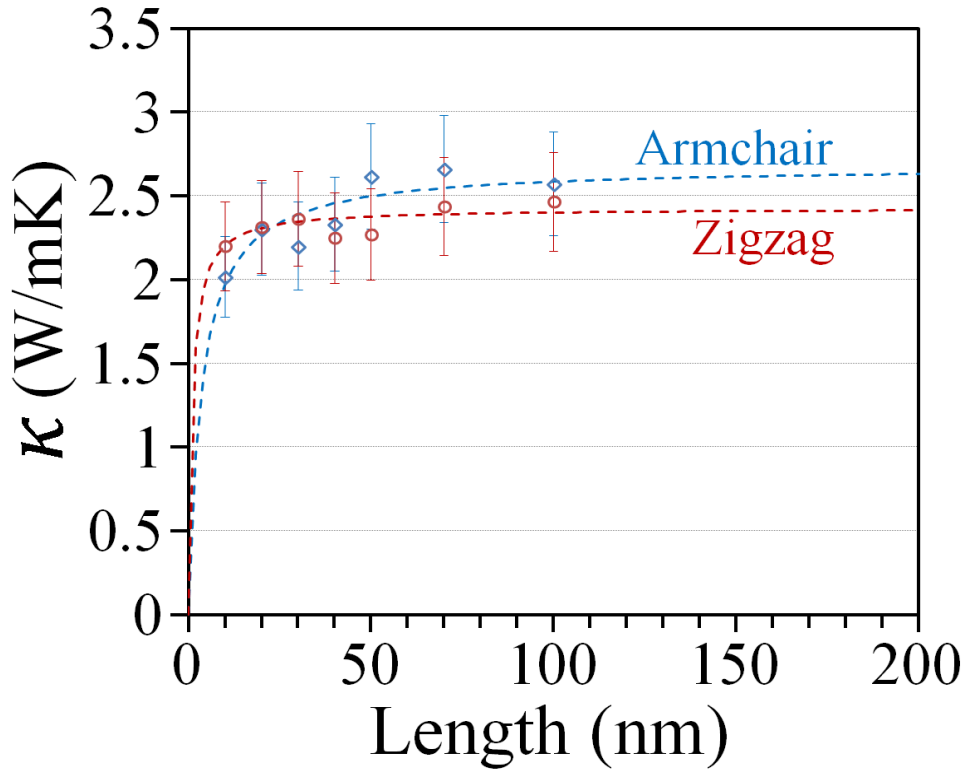

Fig. 5, Non-equilibrium molecular dynamics simulations results for the length effect on the thermal conductivity of single-layer B-graphdiyne along the armchair and zigzag directions at the room temperature.

### 3.3 Thermal conductivity

Thermal conductivity is another important property of a 2D material in the design of nanodevices [75–78]. High thermal conductivity is favourable to avoid overheating issues or application in thermal management systems, whereas a low thermal conductivity is desirable to improve figure of merit of thermoelectric materials. We accordingly
13

also calculate the thermal conductivity of single-layer $C_{12}B_2$, using the classical molecular dynamics simulations. In Fig. 5, the NEMD predictions for the length effect on the thermal conductivity of graphene-like B-graphdiyne along the armchair and zigzag directions are depicted. The results shown in Fig. 5 reveal that the thermal transport along the B-graphdiyne is convincingly isotropic, since the thermal conductivity values are very close along the armchair and zigzag directions. The length effect on the thermal conductivity ($\kappa_L$) of 2D materials usually shows the following relation [79,80]:

$$\frac{1}{\kappa_L} = \frac{1}{\kappa_\infty}\left(1 + \frac{\Lambda}{L}\right) \qquad (6)$$

Here, $L$ is the sample length, $\Lambda$ is the effective phonon mean free path and $\kappa_\infty$ is the thermal conductivity of the sample with infinite length. By fitting the NEMD results using Eq. 6, the lattice thermal conductivity of graphene-like B-graphdiyne along the armchair and zigzag directions were estimated to be 2.68 ±0.3 W/mK and 2.43 ±0.3 W/mK, respectively. The effective phonon mean free along the armchair was also estimated to be 3.6 nm. The phononic thermal conductivity of single-layer B-graphdiyne is by three orders of magnitude smaller than that of the pristine graphene [4,77,81–83].

### 3.4 Electronic and optical properties

To probe the electronic properties of $C_{12}B_2$ monolayer, the band structure, total and partial electronic density of states (EDOS) were calculated. Fig. 6 illustrates the band structure along the high symmetry directions, total and partial EDOS of $C_{12}B_2$ monolayer predicted by PBE method. Our results indicate that the unstrained $C_{12}B_2$ monolayer is a direct band-gap semiconductor at Γ point, which is in a good agreement with the theoretical predictions by Wang *et al.* [36]. The band-gap of free-standing and stress-free $C_{12}B_2$ within the PBE functional was measured to be 0.48 eV. The results shown in Fig. 6 suggest that by applying biaxial loading the electronic band-gap increases (Fig. 6b), whereas applying the uniaxial loading along the armchair (Fig. 6c) and zigzag (Fig. 6d) directions lead to slight decreases in the electronic band-gap. Moreover, it is clear that in unstrained and strained single-layer B-graphdiyne the nearest bands to valence band maximum (VBM) are contributed mainly by the 1C atoms, while above the Fermi energy, nearest bands to conduction band minimum (CBM) are mainly imposed by the connecting B atoms.

Since the PBE functional underestimates the band-gap values, the EDOSs were also computed using the HSE06 hybrid functional. The corresponding band-gap value of unstrained B-graphdiyne within HSE06 was calculated to be 1.15 eV, which is very close to the experimentally measured optical band-gap of 1.1 eV reported by Wang et al. [36]. Taking into consideration that the HSE06 method provides more accurate predictions for the band-gap values, in Fig. 7 we specifically probe the engineering of band-gap in B-graphdiyne, through the mechanical loadings. HSE06 results shown in Fig. 7 suggest that in this monolayer by applying the biaxial loading the electronic band-gap gradually increases. By applying the uniaxial loading along the armchair and zigzag directions the electronic band-gap first slightly increases and then stabilizes and later for larger strain values it starts to decrease. These results highlight the strain tuneable band-gap character in B-graphdiyne monolayer. Moreover, the ultralow thermal conductivity along with direct and narrow band-gap semiconducting electronic character may suggest the graphene-like B-graphdiyne as a promising candidate for the design of novel thermoelectric devices operating at low temperatures.

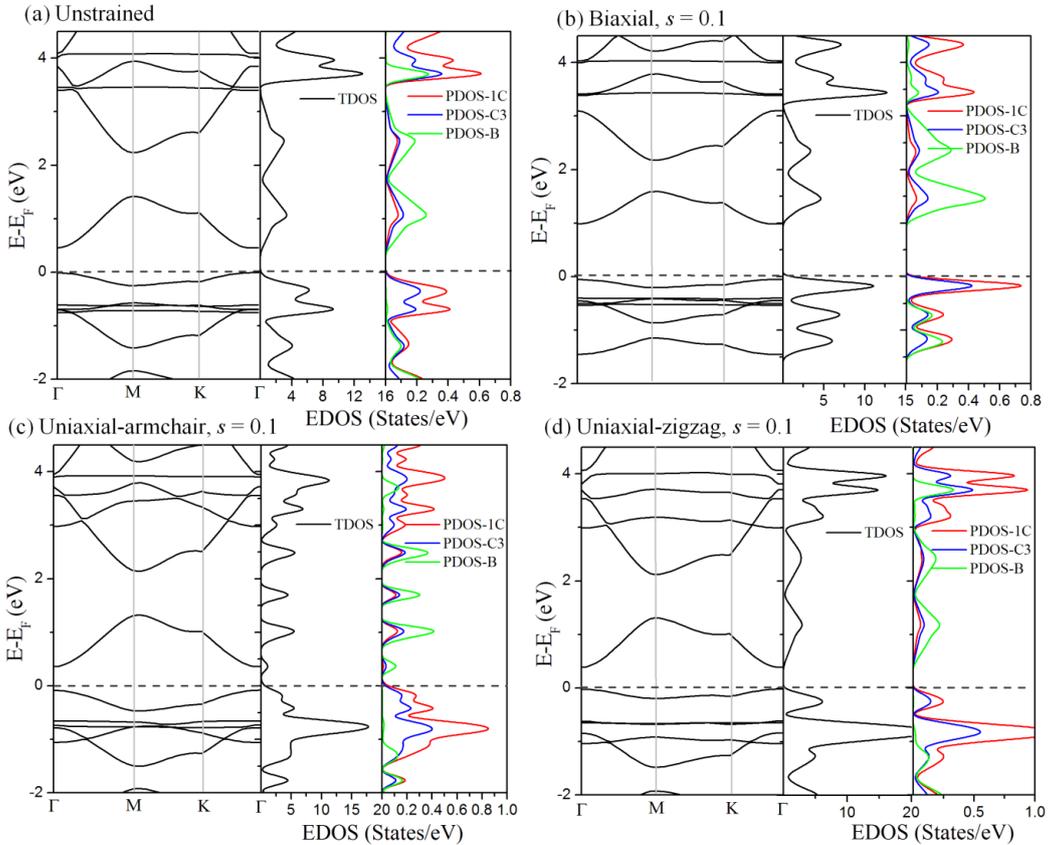

**Fig. 6**, Band structure, total and partial EDOS of unstrained and strained free-standing single-layer $C_{12}B_2$ predicted by the PBE functional. The Fermi energy is aligned to zero.

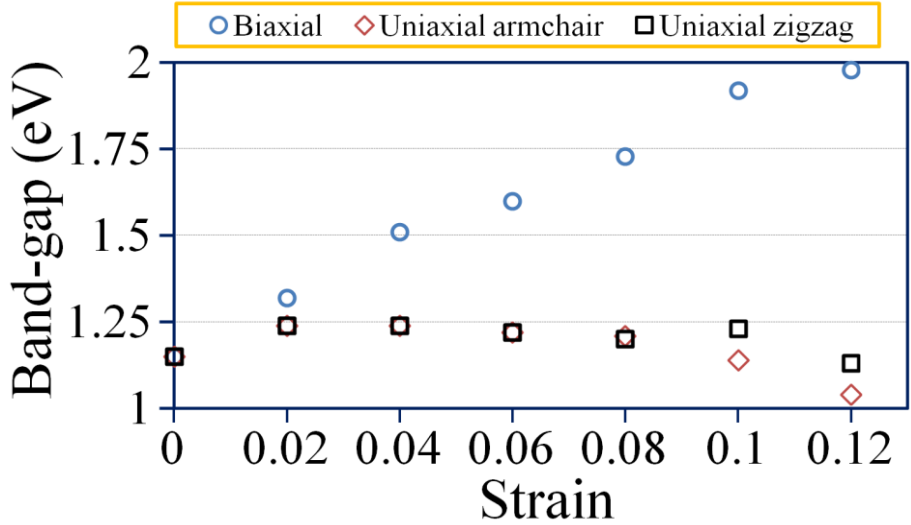

**Fig. 7**, Electronic band-gap of single-layer B-graphdiyne as a function of applied strain predicted by the HSE06 functional.

We next discuss the optical responses of B-graphdiyne monolayer. The imaginary and real parts of the dielectric function of stress-free and strained (with strain of 0.10) $C_{12}B_2$ monolayer for the in-plane ($E//x$ and $E//y$) polarized directions were calculated and the acquired results are illustrated in Fig. 8. As it can be seen the optical spectra for unstrained and biaxial strained systems are isotropic for the light polarizations along the x- and y-axis, whereas it is anisotropic for the uniaxially loaded structures. Moreover, for all cases the optical spectra are highly anisotropic along the in-plane directions. For unstrained monolayer, the first absorption peak of $\text{Im}\varepsilon_{\alpha\beta}(\omega)$ is at 1.14 eV for in-plane polarization, which is in an excellent agreement with the measured experimental optical band-gap of 1.10 eV by Wang et al. [36]. It is conspicuous that by exerting biaxial and uniaxial loading along the zigzag the adsorption edge of $Im\ \varepsilon$ for $E//x$ in the low-frequency regime slightly shifts to higher energies (blue shift) while the shift to lower energies (red shift) is observable for uniaxial loading along the armchair. In contrast, the adsorption edge of $Im\ \varepsilon$ for $E//y$ experiences a red shift for uniaxial loading along the zigzag and a blue shift for uniaxial loading along the armchair. The value of static dielectric constant (the real part of dielectric constant at zero energy, Re $\varepsilon_0$) for the $C_{12}B_2$ monolayer is 3.71 for $E//x$. By exerting biaxial loading the Re $\varepsilon_0$ in all polarizations decreases while by applying the uniaxial loading along the zigzag the static dielectric constant increases along E||x and decreases along E||y. On the other hand for the uniaxial loading along the armchair, the static dielectric constant increases for E||y and decreases along

E||x. Table 1 summarizes the first adsorption peak of *Im ε* and static dielectric constant of unstrained and strained single-layer $C_{12}B_2$.

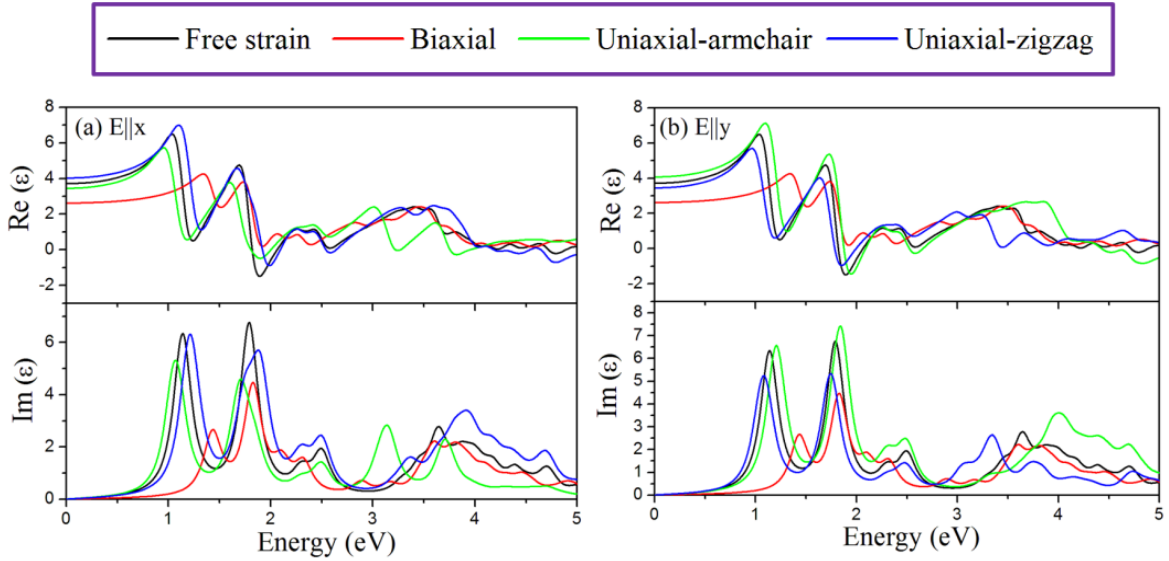

**Fig. 8**, Imaginary and real parts of the dielectric function of single-layer $C_{12}B_2$ with no strain and under the strain level of 0.10 for the in-plane (E||x and E||y) light polarizations, predicted using the PBE plus RPA approach. Insets show *Im ε* in the range of 6-8 eV and *Re ε* near to 0 eV.

Table 1. The first adsorption peak of *Im ε* and static dielectric constant of unstrained and strained free-standing single-layer $C_{12}B_2$.

| Structure | First adsorption peak of *Im ε* (eV) | | Static dielectric constant | |
|---|---|---|---|---|
| | *E//x* | *E//y* | *E//x* | *E//y* |
| Unstrained | 1.14 | 1.14 | 3.71 | 3.71 |
| Biaxial, strain=0.10 | 1.44 | 1.44 | 2.61 | 2.61 |
| Uniaxial-armchair, strain =0.10 | 1.06 | 1.21 | 3.45 | 4.07 |
| Uniaxial-zigzag, strain =0.10 | 1.21 | 1.07 | 4.01 | 3.44 |

The absorption coefficient $α_{ij}(ω)$ for all polarizations are plotted in Fig. 9. The first absorption peak of free strain and strained $C_{12}B_2$ monolayers for in-plane polarization locates in the range of 1.09-1.47 eV which is in the IR and near IR (NIR) range of light. The first absorption peaks become photoactivated at *Γ* wavevector, which is due to electron transition from 1C atoms (valance band) to B atoms (conduction band) (see partial EDOS in Fig. 6). Our results show that the second absorption peaks for the all considered monolayers occur at energy levels around 1.80 eV along the *E//x* and *E//y*, which is desirable for the practical applications in optoelectronic devices in the visible spectral range. The main peak of $α_{ij}(ω)$ for these structures along in-plane polarization is broad and occurs around 3.40–5.20 eV.

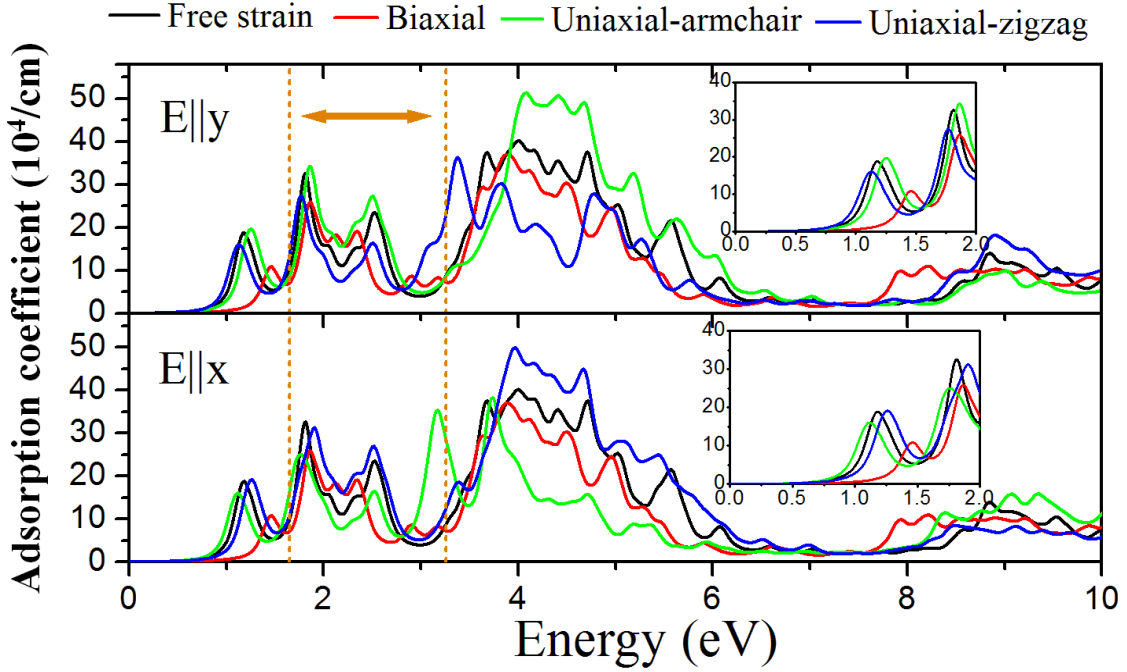

**Fig. 9**, Absorption coefficient of single-layer B-graphdiyne with no strain and under the strain level of 0.10 for the in-plane (E||x and E||y) light polarizations, calculated using the PBE plus RPA approach. Insets show the adsorption coefficient at low frequency range and the visible range of the spectrum is showed by the orange dashed lines.

### 3.5 Application of B-graphdiyne as an anode material

In the rest of this study we elaborately analyze the application prospect of graphene-like B-graphdiyne as an anode material for rechargeable batteries. Appropriate adsorption energy profile of an adatom over the anode material is of a vital importance. We therefore first find the strongest binding sites and their corresponding adsorption energies for Li, Na, Mg and Ca adatoms over the B-graphdiyne. The adsorption energy, $E_{ad}$, in this work was defined by:

$$E_{ad} = E_{TM} - E_T - E_{Ma} \qquad (7)$$

where $E_T$ is the total energy of B-graphdiyne pristine film, $E_{TM}$ is the total energy of the system after metal atoms adsorption and $E_{Ma}$ is the per atom lattice energy of the metal adatoms. The maximum adsorption energies for single Li, Na, Mg and Ca over the B-graphdiyne monolayer were estimated to be, -1.7 eV, -1.75 eV, 0.65 eV and -1.22 eV, respectively. For the all considered adatoms in this work, it was found that they prefer to move inside the monolayer and stay around the B atoms with equal distances from the 1C neighbouring atoms. For the Li and Na adatoms, the most favourable adsorption sites were found to be exactly in the B-graphdiyne lattice plane, whereas for Mg and Ca atoms they show out-of-plane movements. In Fig. 10, the differential charge density plots of the most stable adsorption sites are shown. These results suggest that B-graphdiyne interacts efficiently with the considered elements and accept electron charge densities from them. In this regard and on the basis of Bader charge analysis, from a single Li, Na, Mg and Ca adatoms, charge transfers of 0.994 |e|, 0.993 |e|, 1.467 |e| and 1.469 |e| occur, respectively. The detailed information concerning the most stable adsorption sites of different adatoms over the single-layer B-graphdiyne are provided in the supporting information document. These preliminary results clearly indicate that B-graphdiyne cannot serve as an anode material for Mg ions storage, since the adsorption energy is positive.

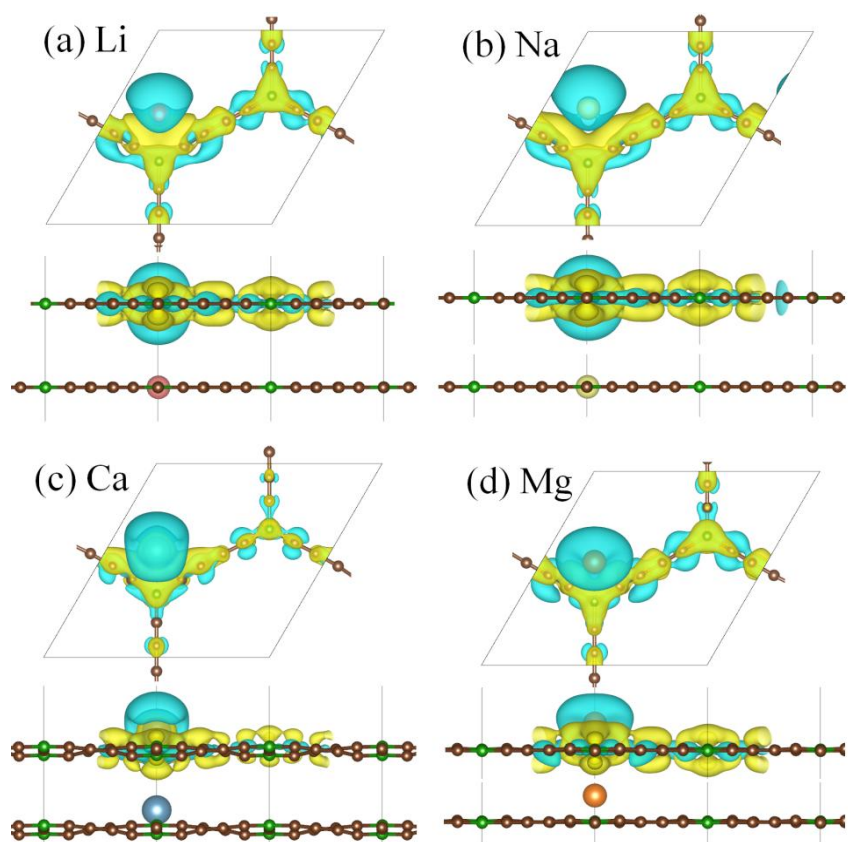

**Fig. 10**, Top and side views of the most favourable adsorption sites for different adatoms in single-layer $C_{12}B_2$. Colour coding illustrate the binding charge transfer due to the adsorption of adatoms over the substrate, in which light-blue shows the charge losses and yellow reveals the charge gains.

To assess the suitability of a material as an anode for rechargeable batteries, the evolution of adsorption energy by increasing the adatoms coverage plays a critical role. It is well-known that the composition and structure of an electrode active material may drastically change by increasing of adatoms coverage. In this work, we therefore gradually and uniformly increased the adatoms coverage. To this aim, we first uniformly placed the metal adatoms on the strongest binding sites. In the B-graphdiyne unit-cell considered in this work, six metal adatoms are required to fill the all most stable binding sites. According to our analysis of adsorption sites, after the most stable site the metal atoms tend to mainly absorb on the top of the B or C atoms. Therefore to simulate the intercalation of adatoms in B-graphdiyne, we positioned the adatoms randomly and uniformly on the both sides of this monolayer. To count for the statistical nature of this problem, we constructed three different systems and after the energy minimization and subsequent single-point energy calculations, the one with the lowest energy was chosen. We then calculated the average adsorption energy, using the following equation:

$$E_{\text{av-ad}} = \frac{(E_{\text{TM}} - n \times E_{Ma} - E_T)}{n} \qquad (8)$$

where $E_{TM}$ is the total energy of B-graphdiyne monolayer with "$n$" metal adatoms adsorbed.

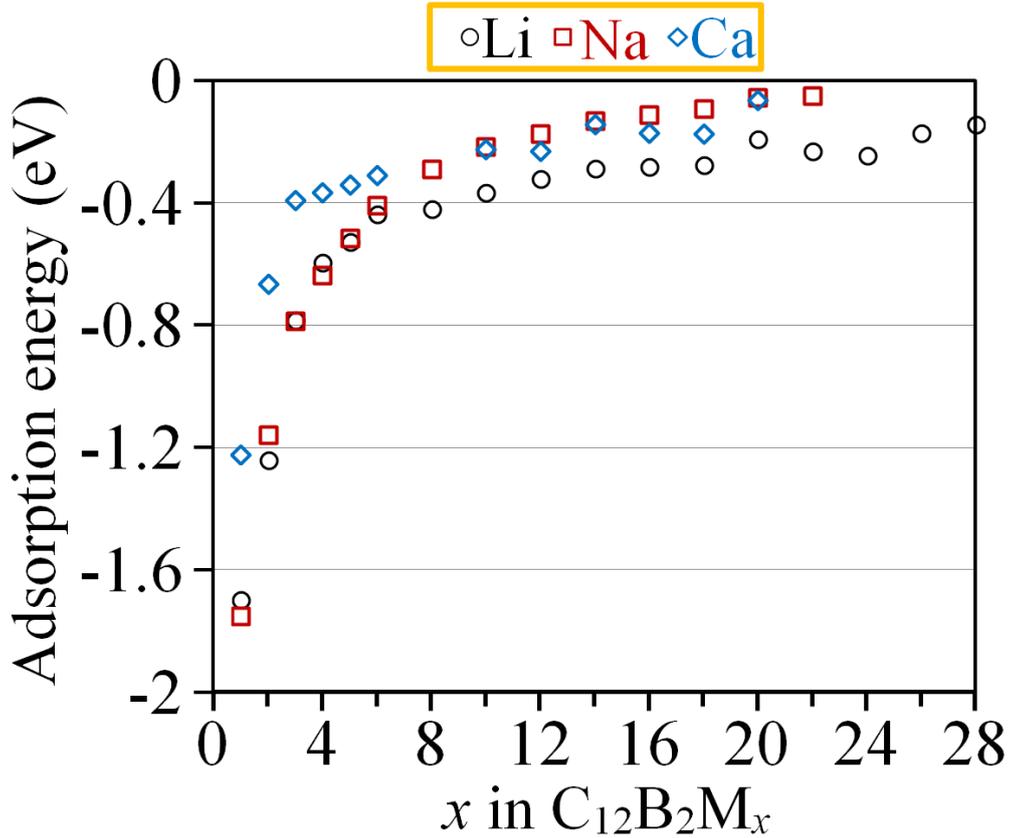

**Fig. 11**, Evolution of average absorption energy as a function of Li, Na and Ca adatoms coverage.

Fig. 11 illustrates the evolution of average adsorption energy of Li, Na and Ca adatoms as a function of coverage. As a general trend for the all considered adatoms, by increasing the coverage the absolute value of adsorption energy decreases. This observation is common since by increasing the adatoms coverage the repulsive forces applied by the previous adsorbed atoms make the bindings of new coming metal atoms more difficult. In the case of Na and Ca atoms, it is obvious that after the adsorption of 20 atoms on the B-graphdiyne monolayer the adsorption energy reaches a negligible value close to the zero. On the other side, for Li atoms after the adsorption of 28 atoms the average adsorption energy is yet negative and far enough from the zero energy. To ensure the structural stability, in Fig. 12 B-graphdiyne monolayer with various adatoms contents are illustrated. As it is clear, this novel 2D structure acts very flexible upon the adatoms adsorption. The maximum bond-lengths in these systems were found to be below 1.7 Å, which is within the safe zone according to our mechanical properties analysis presented earlier. For the Li atoms adsorption over the B-graphdiyne unit-cell, the results shown in Fig. 12 show that when the coverage exceeds from 24 atoms, further Li atoms were pushed out of the surface, forming the second layer of metal Li atoms over the B-graphdiyne surface (see Fig. 12b).

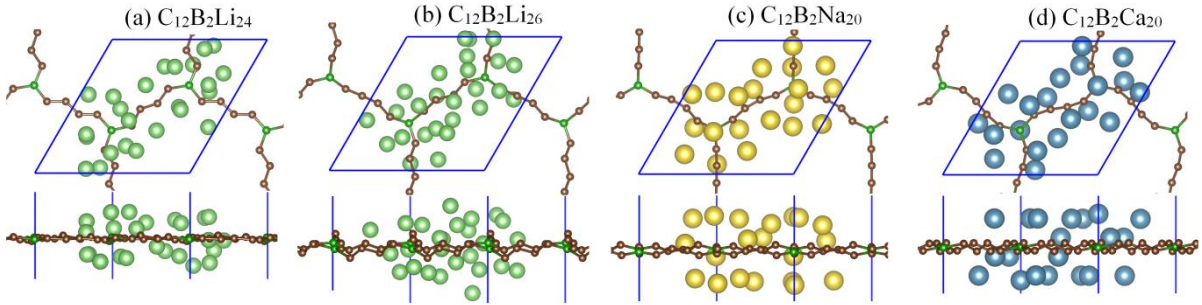

**Fig. 12**, Top and side views of energy minimized single-layer B-graphdiyne with different coverages of Li, Na and Ca atoms. According to our modelling results, B-graphdiyne deform slightly upon the adatoms adsorption. The length criteria for the bonds illustration was chosen to be 1.7 Å.

Another critical factor for the assessment of the performance of a material as an anode is the open-circuit voltage profile. A battery cell output voltage is the difference in the cathode and anode solid phase potentials minus ohmic drops due to contact resistances with the current collectors. As it is clear, in order to increase the cell output voltage, an anode material with lower voltage profile is more desirable. The voltage values close to zero are nevertheless not desirable because that increases the probability of dendrite formation during the adatoms intercalation in anode, which may leads to thermal runaway and other serious damages to the battery. In this work, the average voltages in the coverage range of $x_1 \leq x \leq x_2$ were estimated by the following equation [84,85]:

$$V \approx \frac{(E_{TM_{x_1}} - E_{TM_{x_2}} + (x_2 - x_1)E_{Ma})}{(x_2 - x_1)z} \quad (9)$$

where $E_{TM_{x_1}}$ and $E_{TM_{x_2}}$ are the total energies of the systems with $x_1$ and $x_2$ adsorbed metal adatoms, respectively and $z$ is the valance electrons of an adatom, for Na and Li takes one and for Ca is two. In Fig. 13 the open circuit voltage profiles of the B-graphdiyne monolayer as a function of adatoms coverage are shown. As it is clear, for the all cases after the adsorption of first adatom the voltage drops sharply to below ~0.3 V, which is promising to increase the cell output voltage. Worthy to remind that the negative values of the voltage suggest that foreign adatoms prefer to form metallic clusters instead of adsorption to the active materials. For the Li, Ca and Na atoms storage, after the coverage of, respectively, 22, 16 and 5 atoms over the B-graphdiyne unit-cell the voltage values drop below zero. Based on these results, the specific capacity ($C$) of graphene-like B-graphdiyne can be obtained using; $C = \frac{nZF}{W_{BG}}$, where $n$ is the number of adsorbed metal atoms over the B-graphdiyne unit-cell, $z$ is the number of valance electrons of an adatom type, $F$ is the Faraday constant and $W_{BG}$ is the atomic mass of B-graphdiyne unit-cell. According to the mentioned relation, by considering $n$ equal to 22, 16 and 5 for Li, Ca and Na ions, respectively, it can be concluded that graphene-like B-graphdiyne can exhibit ultrahigh capacities of 808 mAh/g, 5174 mAh/g and 3557 mAh/g for Na, Ca and Li ions storage, respectively.

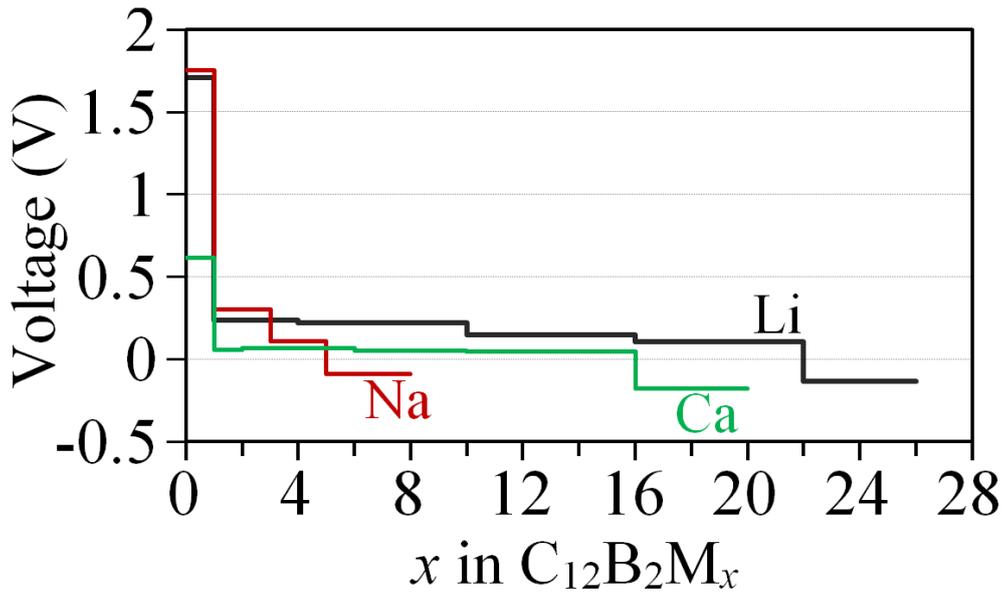

**Fig. 13**, Open circuit voltage profiles as a function of Li, Na and Ca adatoms coverage.

For the application of a material as cathode or anode in rechargeable batteries, presenting the good electronic conductivity is always highly desirable. To reach higher voltage outputs, the internal electronic resistances should be as low as possible. Moreover in order to decrease the risks concerning the overheating issues, like the joule heating, the electronic resistivity of active materials should be low. Worthy to note that experimental measurements confirm the excellent electronic conductivity of B-graphdiyne [36], which is a very appealing character for its practical application as an anode in rechargeable batteries. To probe the effects of adatoms adsorption on the electronic band-gap of B-graphdiyne, we computed the electronic density of states using the HSE06 method. Our results depicted in Fig. 14 confirm that by addition of only a few metal adatoms (three atoms in this case), at the zero state energy (Fermi level) the EDOS is not zero which suggests that the band-gap is closed and the structure becomes metallic.

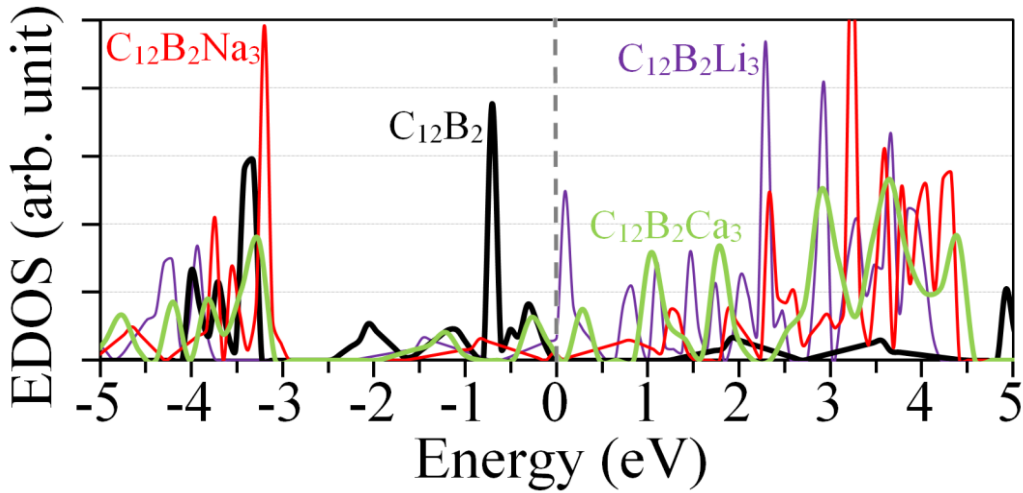

**Fig. 14**, HSE06 results for electronic density of states (EDOS) of the B-graphdiyne monolayer, covered with three adatoms; $C_{12}B_2Li_3$, $C_{12}B_2Na_3$ and $C_{12}B_2Ca_3$, ($C_{12}B_2$ is illustrative of the EDOS for the bare B-graphdiyne). The Fermi level is aligned to zero.

For numerous applications such as those related to mobile communication systems and electric vehicles, fast charging of a battery is among the most critical factors. To explore the suitability of graphene-like B-graphdiyne with respect to the ionic conductivity, we analyzed the diffusion of a single Li adatom using the NEB method and predicted the corresponding energy barriers. To this aim, as a common approach, the diffusion of an adatom between two equivalent most stable adsorption sites was considered.

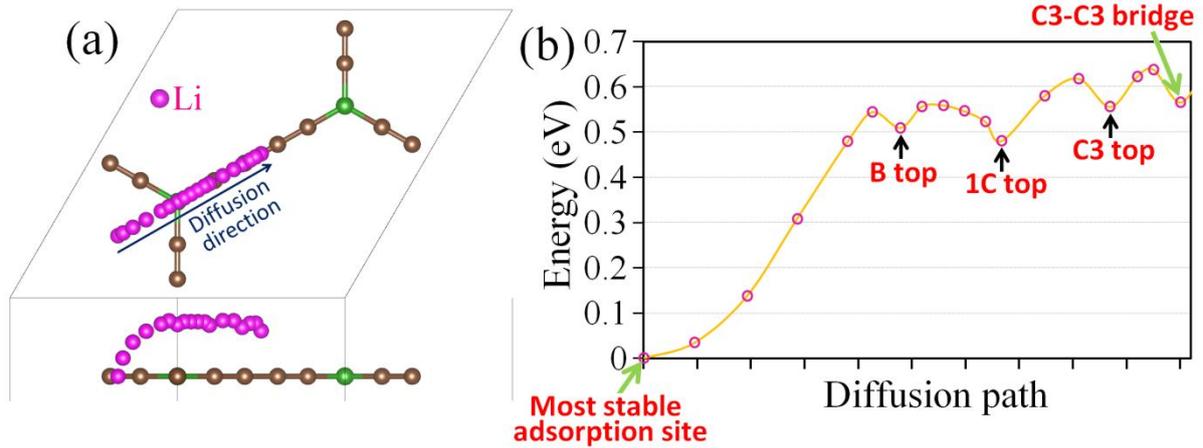

**Fig. 15**, Nudged-elastic band results for the (a) diffusion path and corresponding energy barriers of a Li adatom hopping over the single-layer B-graphdiyne.

Because of the symmetry of graphene-like B-graphdiyne lattice, we analyzed the diffusion of a single Li atom from a most stable adsorption site to the bridging point of C3-C3 carbon bond, which is exactly located at the center of the B-graphdiyne unit-cell lattice. The considered diffusion path for a single Li atom is illustrated in Fig. 15a. In Fig. 15a the NEB images for the Li atom diffusion process over the single-layer B-graphdiyne are shown. We note that in our NEB analysis all atoms in the systems were allowed to move and accordingly rearrange their atomic positions upon the Li atoms movements. Nevertheless, in Fig. 15a in order to simplify the understanding of this illustration, the positions of B and C atoms were fixed. In Fig. 15b the corresponding relative energies for every NEB image are depicted. Interestingly, the Li adatom first jumps over the B atoms passing an energy barrier of 0.54 eV. The diffusion is followed by jumping over the 1C and C3 carbon atoms, overtaking energy barriers of 0.05 eV and 0.14 eV, respectively. The considered diffusion path is then accomplished by reaching the bridging point of C3-C3 bond passing a small energy barrier of 0.08 eV. As it is clear, in this system the diffusion mechanism is not uniform. The maximum energy barrier for the diffusions of an Li ion over the carbon chains in B-graphdiyne is 0.14 eV, which is considerably lower than that over the graphene ~0.37 eV [86]. Worthy to remind that according to the Arrhenius equation, the diffusion rate of adatoms shows an exponential depence on the energy barrier and therefore even a small reduction in the energy barrier can lead to considerable changes in the diffusion rate. Based on these results, for the charging process the Li ions can easily fill the B-graphdiyne anode material through fast diffusions over the carbon chains. In this case, for the Li ions located at the top of the B atoms only a very small energy barrier of 0.035 eV has to be passed to fill the most stable adsorption sites. Moreover, the porous structure of B-graphdiyne provides excellent conditions for the diffusion of Li ions through different atomic layers. Nevertheless, the high energy barrier of 0.54 eV for the hopping of Li atoms form the most stable adsorption sites over the B atoms reveal the low diffusion rate of Li atoms over the B-graphdiyne for the discharging process. Our results for the diffusion energy barriers highlight the outstanding prospect of B-graphdiyne to achieve fast charging rates, very decisive for success in critical technologies like mobile electronics and electric vehicles.

## 4. Conclusion

In a most recent experimental advance, B-graphdiyne, a novel graphene-like 2D structure was synthesized. This exciting experimental achievement introduces the B-graphdiyne, as a new class of 2D materials with low-density and porous atomic lattices made only from carbon and boron atoms. Motivated by the recent experimental advance in the fabrication of graphene-like B-graphdiyne with an empirical formula of $C_{12}B_2$, we conducted extensive first-principles DFT and classical molecular dynamics simulations to explore the mechanical properties, thermal conductivity and stability, electronic and optical responses of free-standing and single-layer B-graphdiyne. Mechanical properties were studied by performing the uniaxial tensile simulations within the DFT method. The ultimate tensile strength of single-layer $C_{12}B_2$ along the armchair and zigzag directions were predicted to be 15 N/m and 9.6 N/m, respectively. It was found that the B-graphdiyne monolayer when stretched along the armchair direction can exhibit ultrahigh stretchability, outperforming other known 2D pristine materials. Ab initio molecular dynamics simulations confirm the outstanding thermal stability of graphene-like B-graphdiyne monolayer, which can stay intact at high temperatures up to 2500 K. It was predicted that the mechanical and thermal failures in the B-graphdiyne nanomembranes initiate by the breakage of the B-C bonds. According to our classical non-equilibrium molecular dynamics simulation results, the thermal conductivity of B-graphdiyne monolayer at the room temperature was found to be ultralow, ~2.5 W/mK.

The first-principles results confirm that $C_{12}B_2$ monolayer is a direct band-gap semiconductor. The electronic band-gap of graphene-like B-graphdiyne was predicted to be 1.15 eV within the HSE06 method. We employed the RPA to calculate the complex dielectric function of this monolayer. The first absorption peak was found to happen at 1.14 eV for in-plane polarizations. The first absorption peak of the graphene-like B-graphdiyne monolayer for in-plane polarization locates in the IR and near IR (NIR) range of light. Our results for strained structures confirm that the electronic and optical properties of single-layer $C_{12}B_2$ can be tuned by the mechanical straining. The ultralow thermal conductivity along with the direct band-gap semiconducting electronic character may also suggest the B-graphdiyne nanomembranes as promising candidates for the design of novel carbon based thermoelectric devices.

We finally explored the potential application of graphene-like B-graphdiyne as an anode material for Li, Na, Ca and Mg ions storage. It was found that this novel 2D structure is suitable for Li, Na and Ca ions storage. The results for the open circuit voltage profiles reveal the promising performance of B-graphdiyne nanomembranes to reach high output voltages. According to our first-principles results, graphene-like B-graphdiyne was predicted to yield ultrahigh charge capacities of 808 mAh/g, 5174 mAh/g and 3557 mAh/g for Na, Ca and Li ions storage, respectively. Nudged elastic band results for the diffusion of a single Li atom suggest the B-graphdiyne as a promising anode material to design batteries suitable for fast charging rates.

### Acknowledgment

B. M. and T. R. greatly acknowledge the financial support by European Research Council for COMBAT project (Grant number 615132).

### References

(1) Novoselov, K. S.; Geim, A. K.; Morozov, S. V; Jiang, D.; Zhang, Y.; Dubonos, S. V; Grigorieva, I. V; Firsov, A. A. Electric Field Effect in Atomically Thin Carbon Films. *Science* **2004**, *306*, 666–669.

(2) Geim, A. K.; Novoselov, K. S. The Rise of Graphene. *Nat. Mater.* **2007**, *6*, 183–191.

(3) Lee, C.; Wei, X.; Kysar, J. W.; Hone, J. Measurement of the Elastic Properties and Intrinsic Strength of Monolayer Graphene. *Science (80-. ).* **2008**, *321*, 385–388.

(4) Ghosh, S.; Bao, W.; Nika, D. L.; Subrina, S.; Pokatilov, E. P.; Lau, C. N.; Balandin, A. a. Dimensional Crossover of Thermal Transport in Few-Layer Graphene. *Nat. Mater.* **2010**, *9*, 555–558.

(5) Thomas, A.; Fischer, A.; Goettmann, F.; Antonietti, M.; Müller, J.-O.; Schlögl, R.; Carlsson, J. M. Graphitic Carbon Nitride Materials: Variation of Structure and Morphology and Their Use as Metal-Free Catalysts. *J. Mater. Chem.* **2008**, *18*, 4893.

(6) Algara-Siller, G.; Severin, N.; Chong, S. Y.; Björkman, T.; Palgrave, R. G.; Laybourn, A.; Antonietti, M.; Khimyak, Y. Z.; Krasheninnikov, A. V.; Rabe, J. P.; *et al.* Triazine-Based Graphitic Carbon Nitride: A Two-Dimensional Semiconductor. *Angew.*

*Chemie - Int. Ed.* **2014**, *53*, 7450–7455.

(7) Wang, X.; Maeda, K.; Thomas, A.; Takanabe, K.; Xin, G.; Carlsson, J. M.; Domen, K.; Antonietti, M. A Metal-Free Polymeric Photocatalyst for Hydrogen Production from Water under Visible Light. *Nat. Mater.* **2009**, *8*, 76–80.

(8) Zheng, Y.; Jiao, Y.; Chen, J.; Liu, J.; Liang, J.; Du, A.; Zhang, W.; Zhu, Z.; Smith, S. C.; Jaroniec, M.; *et al.* Nanoporous Graphitic-C 3 N 4 @Carbon Metal-Free Electrocatalysts for Highly Efficient Oxygen Reduction. *J. Am. Chem. Soc.* **2011**, *133*, 20116–20119.

(9) Lyth, S. M.; Nabae, Y.; Islam, N. M.; Kuroki, S.; Kakimoto, M.; Miyata, S. Electrochemical Oxygen Reduction Activity of Carbon Nitride Supported on Carbon Black. *J. Electrochem. Soc.* **2011**, *158*, B194–B201.

(10) Lyth, S. M.; Nabae, Y.; Moriya, S.; Kuroki, S.; Kakimoto, M. A.; Ozaki, J. I.; Miyata, S. Carbon Nitride as a Nonprecious Catalyst for Electrochemical Oxygen Reduction. *J. Phys. Chem. C* **2009**, *113*, 20148–20151.

(11) Zhu, J.; Xiao, P.; Li, H.; Carabineiro, S. a C. Graphitic Carbon Nitride: Synthesis, Properties, and Applications in Catalysis. *ACS Appl. Mater. Interfaces* **2014**, *6*, 16449–16465.

(12) de Sousa, J. M.; Botari, T.; Perim, E.; Bizao, R. A.; Galvao, D. S. Mechanical and Structural Properties of Graphene-like Carbon Nitride Sheets. *RSC Adv.* **2016**, *6*, 76915–76921.

(13) Mahmood, J.; Lee, E. K.; Jung, M.; Shin, D.; Jeon, I.-Y.; Jung, S.-M.; Choi, H.-J.; Seo, J.-M.; Bae, S.-Y.; Sohn, S.-D.; *et al.* Nitrogenated Holey Two-Dimensional Structures. *Nat. Commun.* **2015**, *6*, 6486.

(14) Mahmood, J.; Lee, E. K.; Jung, M.; Shin, D.; Choi, H.-J.; Seo, J.-M.; Jung, S.-M.; Kim, D.; Li, F.; Lah, M. S.; *et al.* Two-Dimensional Polyaniline (C3N) from Carbonized Organic Single Crystals in Solid State. *Proc. Natl. Acad. Sci.* **2016**, *113*, 7414–7419.

(15) Shi, L.-B.; Zhang, Y.-Y.; Xiu, X.-M.; Dong, H.-K. Structural, Electronic and Adsorptive Characteristics of Phosphorated Holey Graphene (PHG): First Principles Calculations. *Diam. Relat. Mater.* **2018**, *82*, 102–108.

(16) Shi, L.-B.; Zhang, Y.-Y.; Xiu, X.-M.; Dong, H.-K. Structural Characteristics and Strain Behaviors of Two-Dimensional C3N : First Principles Calculations. *Carbon N. Y.* **2018**, *134*, 103–111.

(17) Makaremi, M.; Grixti, S.; Butler, K. T.; Ozin, G. A.; Singh, C. V. Band Engineering of Carbon Nitride Monolayers by N-Type, P-Type, and Isoelectronic Doping for Photocatalytic Applications. *ACS Appl. Mater. Interfaces* **2018**, *10*, 11143–11151.

(18) Gao, Y.; Wang, H.; Sun, M.; Ding, Y.; Zhang, L.; Li, Q. First-Principles Study of Intrinsic Phononic Thermal Transport in Monolayer C3N. *Phys. E Low-dimensional*

*Syst. Nanostructures* **2018**, *99*, 194–201.

(19) Wang, X.; Li, Q.; Wang, H.; Gao, Y.; Hou, J.; Shao, J. Anisotropic Carrier Mobility in Single- and Bi-Layer C3N Sheets. *Phys. B Condens. Matter* **2018**, *537*, 314–319.

(20) Baughman, R. H.; Eckhardt, H.; Kertesz, M. Structure-Property Predictions for New Planar Forms of Carbon: Layered Phases Containing sp$^{2}$ and Sp Atoms. *J. Chem. Phys.* **1987**, *87*, 6687.

(21) Jia, Z.; Zuo, Z.; Yi, Y.; Liu, H.; Li, D.; Li, Y.; Li, Y. *Low Temperature, Atmospheric Pressure for Synthesis of a New Carbon Ene-Yne and Application in Li Storage*; 2017; Vol. 33.

(22) Matsuoka, R.; Sakamoto, R.; Hoshiko, K.; Sasaki, S.; Masunaga, H.; Nagashio, K.; Nishihara, H. Crystalline Graphdiyne Nanosheets Produced at a Gas/Liquid or Liquid/Liquid Interface. *J. Am. Chem. Soc.* **2017**, *139*, 3145–3152.

(23) Makaremi, M.; Mortazavi, B.; Singh, C. V. Carbon Ene-Yne Graphyne Monolayer as an Outstanding Anode Material for Li/Na Ion Batteries. *Appl. Mater. Today* **2018**, *10*.

(24) Hwang, H. J.; Koo, J.; Park, M.; Park, N.; Kwon, Y.; Lee, H. Multilayer Graphynes for Lithium Ion Battery Anode. *J. Phys. Chem. C* **2013**, *117*, 6919–6923.

(25) Bartolomei, M.; Carmona-Novillo, E.; Giorgi, G. First Principles Investigation of Hydrogen Physical Adsorption on Graphynes' Layers. *Carbon N. Y.* **2015**, *95*, 1076–1081.

(26) Autreto, P. A. S.; De Sousa, J. M.; Galvao, D. S. Site-Dependent Hydrogenation on Graphdiyne. *Carbon N. Y.* **2014**, *77*, 829–834.

(27) Hwang, H. J.; Kwon, Y.; Lee, H. Thermodynamically Stable Calcium-Decorated Graphyne as a Hydrogen Storage Medium. *J. Phys. Chem. C* **2012**, *116*, 20220–20224.

(28) Yao, Y.; Jin, Z.; Chen, Y.; Gao, Z.; Yan, J.; Liu, H.; Wang, J.; Li, Y.; Liu, S. (Frank). Graphdiyne-WS22D-Nanohybrid Electrocatalysts for High-Performance Hydrogen Evolution Reaction. *Carbon N. Y.* **2018**, *129*, 228–235.

(29) Lin, Z. Z. Graphdiyne as a Promising Substrate for Stabilizing Pt Nanoparticle Catalyst. *Carbon N. Y.* **2015**, *86*, 301–309.

(30) Sun, L.; Jiang, P. H.; Liu, H. J.; Fan, D. D.; Liang, J. H.; Wei, J.; Cheng, L.; Zhang, J.; Shi, J. Graphdiyne: A Two-Dimensional Thermoelectric Material with High Figure of Merit. *Carbon N. Y.* **2015**, *90*, 255–259.

(31) Wang, X. M.; Lu, S. S. Thermoelectric Transport in Graphyne Nanotubes. *J. Phys. Chem. C* **2013**, *117*, 19740–19745.

(32) Ketabi, N.; Tolhurst, T. M.; Leedahl, B.; Liu, H.; Li, Y.; Moewes, A. How Functional Groups Change the Electronic Structure of Graphdiyne: Theory and Experiment. *Carbon N. Y.* **2017**, *123*, 1–6.

(33) Ruiz-Puigdollers, A.; Gamallo, P. DFT Study of the Role of N- and B-Doping on

Structural, Elastic and Electronic Properties of α -, β - and γ -Graphyne. *Carbon N. Y.* **2017**, *114*, 301–310.

(34) Hu, M.; Pan, Y.; Luo, K.; He, J.; Yu, D.; Xu, B. Three Dimensional Graphdiyne Polymers with Tunable Band Gaps. *Carbon N. Y.* **2015**, *91*, 518–526.

(35) Li, Z.; Smeu, M.; Rives, A.; Maraval, V.; Chauvin, R.; Ratner, M. A.; Borguet, E. Towards Graphyne Molecular Electronics. *Nat. Commun.* **2015**, *6*, 6321.

(36) Wang, N.; Li, X.; Tu, Z.; Zhao, F.; He, J.; Guan, Z.; Huang, C.; Yi, Y.; Li, Y. Synthesis, Electronic Structure of Boron-Graphdiyne with an Sp-Hybridized Carbon Skeleton and Its Application in Sodium Storage. *Angew. Chemie* **2018**.

(37) Shi, L.; Zhao, T. Recent Advances in Inorganic 2D Materials and Their Applications in Lithium and Sodium Batteries. *J. Mater. Chem. A* **2017**, *5*, 3735–3758.

(38) Samad, A.; Noor-A-Alam, M.; Shin, Y.-H. First Principles Study of a SnS2/graphene Heterostructure: A Promising Anode Material for Rechargeable Na Ion Batteries. *J. Mater. Chem. A* **2016**, *4*, 14316–14323.

(39) Samad, A.; Shafique, A.; Kim, H. J.; Shin, Y.-H. Superionic and Electronic Conductivity in Monolayer W2C: Ab Initio Predictions. *J. Mater. Chem. A* **2017**, *5*, 11094–11099.

(40) Pomerantseva, E.; Gogotsi, Y. Two-Dimensional Heterostructures for Energy Storage. *Nat. Energy* **2017**, *2*, 17089.

(41) Zhang, X.; Hou, L.; Ciesielski, A.; Samorì, P. 2D Materials Beyond Graphene for High-Performance Energy Storage Applications. *Adv. Energy Mater.* **2016**, *6*, n/a--n/a.

(42) Sibari, A.; El Marjaoui, A.; Lakhal, M.; Kerrami, Z.; Kara, A.; Benaissa, M.; Ennaoui, A.; Hamedoun, M.; Benyoussef, A.; Mounkachi, O. Phosphorene as a Promising Anode Material for (Li/Na/Mg)-Ion Batteries: A First-Principle Study. *Sol. Energy Mater. Sol. Cells* **2017**.

(43) Grixti, S.; Mukherjee, S.; Singh, C. V. Two- dimensional Boron as an Impressive Lithium- sulphur Battery Cathode Material. *Energy Storage Mater.* **2018**, *13*, 80–87.

(44) Wang, Y.; Song, N.; Song, X.; Zhang, T.; Zhang, Q.; Li, M. Metallic VO2 Monolayer as an Anode Material for Li, Na, K, Mg or Ca Ion Storage: A First-Principle Study. *RSC Adv.* **2018**, *8*, 10848–10854.

(45) Yu, S.; Rao, Y.-C.; Li, S.-F.; Duan, X.-M. Net W Monolayer: A High-Performance Electrode Material for Li-Ion Batteries. *Appl. Phys. Lett.* **2018**, *112*, 53903.

(46) Kresse, G. From Ultrasoft Pseudopotentials to the Projector Augmented-Wave Method. *Phys. Rev. B* **1999**, *59*, 1758–1775.

(47) Kresse, G.; Furthm??ller, J. Efficiency of Ab-Initio Total Energy Calculations for Metals and Semiconductors Using a Plane-Wave Basis Set. *Comput. Mater. Sci.* **1996**, *6*, 15–50.

(48) Kresse, G.; Furthmüller, J. Efficient Iterative Schemes for Ab Initio Total-Energy Calculations Using a Plane-Wave Basis Set. *Phys. Rev. B* **1996**, *54*, 11169–11186.

(49) Perdew, J.; Burke, K.; Ernzerhof, M. Generalized Gradient Approximation Made Simple. *Phys. Rev. Lett.* **1996**, *77*, 3865–3868.

(50) Blöchl, P. E. Projector Augmented-Wave Method. *Phys. Rev. B* **1994**, *50*, 17953–17979.

(51) Momma, K.; Izumi, F. VESTA 3 for Three-Dimensional Visualization of Crystal, Volumetric and Morphology Data. *J. Appl. Crystallogr.* **2011**, *44*, 1272–1276.

(52) Monkhorst, H.; Pack, J. Special Points for Brillouin Zone Integrations. *Phys. Rev. B* **1976**, *13*, 5188–5192.

(53) Scuseria, A. V. K. and O. A. V. and A. F. I. and G. E. Influence of the Exchange Screening Parameter on the Performance of Screened Hybrid Functionals. *J. Chem. Phys.* **2006**, *125*, 224106.

(54) Hohenberg, P.; Kohn, W. The Inhomogeneous Electron Gas. *Phys. Rev.* **1964**, *136*, B864.

(55) Wooten, F. *Optical Properties of Solids*; Academic press, 2013.

(56) Shahrokhi, M.; Leonard, C. Quasi-Particle Energies and Optical Excitations of Wurtzite BeO and Its Nanosheet. *J. Alloys Compd.* **2016**, *682*, 254–262.

(57) Shahrokhi, M. Quasi-Particle Energies and Optical Excitations of ZnS Monolayer Honeycomb Structure. *Appl. Surf. Sci.* **2016**, *390*, 377–384.

(58) Plimpton, S. Fast Parallel Algorithms for Short-Range Molecular Dynamics. *J. Comput. Phys.* **1995**, *117*, 1–19.

(59) Tersoff, J. Modeling Solid-State Chemistry: Interatomic Potentials for Multicomponent Systems. *Phys. Rev. B* **1989**, *39*, 5566–5568.

(60) L. Lindsay, D. A. B. Optimized Tersoff and Brenner Empirical Potential Parameters for Lattice Dynamics and Phonon Thermal Transport in Carbon Nanotubes and Graphene. *Phys. Rev. B - Condens. Matter Mater. Phys.* **2010**, *82*, 205441.

(61) KInacI, A.; Haskins, J. B.; Sevik, C.; ÇağIn, T. Thermal Conductivity of BN-C Nanostructures. *Phys. Rev. B - Condens. Matter Mater. Phys.* **2012**, *86*.

(62) Bazrafshan, S.; Rajabpour, A. Engineering of Thermal Transport in Graphene Using Grain Size, Strain, Nitrogen and Boron Doping; a Multiscale Modeling. *Int. J. Heat Mass Transf.* **2018**, *123*, 534–543.

(63) Hong, Y.; Zhang, J.; Zeng, X. C. Monolayer and Bilayer Polyaniline C3N: Two-Dimensional Semiconductors with High Thermal Conductivity. *Nanoscale* **2018**, *10*, 4301–4310.

(64) Felix, I. M.; Pereira, L. F. C. Thermal Conductivity of Graphene-hBN Superlattice Ribbons. *Sci. Rep.* **2018**, *8*, 2737.

(65) Dong, Y.; Meng, M.; Groves, M. M.; Zhang, C.; Lin, J. Thermal Conductivities of

Two-Dimensional Graphitic Carbon Nitrides by Molecule Dynamics Simulation. *Int. J. Heat Mass Transf.* **2018**, *123*, 738–746.

(66) Grimme, S. Semiempirical GGA-Type Density Functional Constructed with a Long-Range Dispersion Correction. *J. Comput. Chem.* **2006**, *27*, 1787–1799.

(67) Blöchl, P. E.; Jepsen, O.; Andersen, O. K. Improved Tetrahedron Method for Brillouin-Zone Integrations. *Phys. Rev. B* **1994**, *49*, 16223–16233.

(68) Silvi, B.; Savin, A. Classification of Chemical-Bonds Based on Topological Analysis of Electron Localization Functions. *Nature* **1994**, *371*, 683–686.

(69) Shahrokhi, M.; Leonard, C. Tuning the Band Gap and Optical Spectra of Silicon-Doped Graphene: Many-Body Effects and Excitonic States. *J. Alloys Compd.* **2017**, *693*, 1185–1196.

(70) Liu, F.; Ming, P.; Li, J. Ab Initio Calculation of Ideal Strength and Phonon Instability of Graphene under Tension. *Phys. Rev. B - Condens. Matter Mater. Phys.* **2007**, *76*.

(71) Peng, Q.; Ji, W.; De, S. Mechanical Properties of the Hexagonal Boron Nitride Monolayer: Ab Initio Study. *Comput. Mater. Sci.* **2012**, *56*, 11–17.

(72) Sun, H.; Mukherjee, S.; Daly, M.; Krishnan, A.; Karigerasi, M. H.; Singh, C. V. New Insights into the Structure-Nonlinear Mechanical Property Relations for Graphene Allotropes. *Carbon N. Y.* **2016**, *110*, 443–457.

(73) Mortazavi, B.; Shahrokhi, M.; Rabczuk, T.; Pereira, L. F. C. Electronic, Optical and Thermal Properties of Highly Stretchable 2D Carbon Ene-Yne Graphyne. *Carbon N. Y.* **2017**, *123*, 344–353.

(74) Peng, Q.; Ji, W.; De, S. Mechanical Properties of Graphyne Monolayers: A First-Principles Study. *Phys. Chem. Chem. Phys.* **2012**, *14*, 13385.

(75) Mortazavi, B.; Rabczuk, T. Multiscale Modelling of Heat Conduction in All-MoS2 Single-Layer Heterostructures. *RSC Adv.* **2017**, *7*, 11135–11141.

(76) Nika, D. L.; Balandin, A. A. Two-Dimensional Phonon Transport in Graphene. *J. Phys. Condens. Matter* **2012**, *24*, 233203.

(77) Balandin, A. A. Thermal Properties of Graphene and Nanostructured Carbon Materials. *Nat. Mater.* **2011**, *10*, 569–581.

(78) Shahil, K. M. F.; Balandin, A. A. Thermal Properties of Graphene and Multilayer Graphene: Applications in Thermal Interface Materials. *Solid State Communications*, 2012, *152*, 1331–1340.

(79) Schelling, P. K.; Phillpot, S. R.; Keblinski, P. Comparison of Atomic-Level Simulation Methods for Computing Thermal Conductivity. *Phys. Rev. B* **2002**, *65*, 1–12.

(80) Zhang, X.; Xie, H.; Hu, M.; Bao, H.; Yue, S.; Qin, G.; Su, G. Thermal Conductivity of Silicene Calculated Using an Optimized Stillinger-Weber Potential. *Phys. Rev. B - Condens. Matter Mater. Phys.* **2014**, *89*.

(81) Balandin, A. A.; Ghosh, S.; Bao, W.; Calizo, I.; Teweldebrhan, D.; Miao, F.; Lau, C. N. Superior Thermal Conductivity of Single-Layer Graphene. *Nano Lett.* **2008**, *8*, 902–907.

(82) Rajabpour, A.; Vaez Allaei, S. M.; Kowsary, F. Interface Thermal Resistance and Thermal Rectification in Hybrid Graphene-Graphane Nanoribbons: A Nonequilibrium Molecular Dynamics Study. *Appl. Phys. Lett.* **2011**, *99*.

(83) Rajabpour, A.; Vaez Allaei, S. M. Tuning Thermal Conductivity of Bilayer Graphene by Inter-Layer sp3 Bonding: A Molecular Dynamics Study. *Appl. Phys. Lett.* **2012**, *101*, 53115.

(84) Zhou, F.; Cococcioni, M.; Marianetti, C. A.; Morgan, D.; Ceder, G. First-Principles Prediction of Redox Potentials in Transition-Metal Compounds with LDA + U. *Phys. Rev. B - Condens. Matter Mater. Phys.* **2004**, *70*, 1–8.

(85) Aydinol, M.; Kohan, a.; Ceder, G.; Cho, K.; Joannopoulos, J. Ab Initio Study of Lithium Intercalation in Metal Oxides and Metal Dichalcogenides. *Phys. Rev. B* **1997**, *56*, 1354–1365.

(86) Pollak, E.; Geng, B.; Jeon, K. J.; Lucas, I. T.; Richardson, T. J.; Wang, F.; Kostecki, R. The Interaction of Li+ with Single-Layer and Few-Layer Graphene. *Nano Lett.* **2010**, *10*, 3386–3388.

# Supporting Information

# Boron-graphdiyne: superstretchable semiconductor with low thermal conductivity and ultrahigh capacity for Li, Na and Ca ions storage

Bohayra Mortazavi[*,1], Masoud Shahrokhi[2], Xiaoying Zhuang[3] and Timon Rabczuk[4]

[1]*Institute of Structural Mechanics, Bauhaus-Universität Weimar, Marienstr. 15, D-99423 Weimar, Germany.*
[2]*Institute of Chemical Research of Catalonia, ICIQ, The Barcelona Institute of Science and Technology, Av. Països Catalans 16, ES-43007 Tarragona, Spain.*
[3]*Institut für Kontinuumsmechanik, Gottfried Wilhelm Leibniz Universität Hannover, Appelstrasse 11, 30167 Hannover, Germany.*
[4]*College of Civil Engineering, Department of Geotechnical Engineering, Tongji University, Shanghai, China.*

*E-mail: bohayra.mortazavi@gmail.com

1. Atomic structure of boron-graphdiyne unit-cell

2. Most stable adsorption sites for Li, Na, Mg and Ca atoms over B-graphdiyne.

## 1. Atomic structure of B-graphdiyne unit-cell

```
C12B2
   1.00000000000000
     11.8467556014048103    0.0000000000000000    0.0000000000000000
      5.9233778007024069   10.2595913032290600    0.0000000000000000
      0.0000000000000000    0.0000000000000000   20.0000000000000000
   C    B
   12    2
Direct
  0.4070313359871207   0.1859316676351952   0.5000000000000000
  0.4672120785149971   0.0655705418981327   0.5000000000000000
  0.4070313102567411   0.4070303586203110   0.5000000000000000
  0.4672113650511491   0.4672112185712862   0.5000000000000000
  0.5327836106014843   0.5327839635983835   0.5000000000000000
  0.5929637664429919   0.5929648549585735   0.5000000000000000
  0.5929638727348205   0.8140640368300751   0.5000000000000000
  0.5327845410920844   0.9344251783120043   0.5000000000000000
  0.1859315430662534   0.4070331135330179   0.5000000000000000
  0.0655704174789662   0.4672136888685117   0.5000000000000000
  0.8140642497539758   0.5929627139400750   0.5000000000000000
  0.9344251513708954   0.5327831630738729   0.5000000000000000
  0.3333313296681197   0.3333314582972733   0.5000000000000000
  0.6666641611622097   0.6666642439564683   0.5000000000000000
```

# 2. Most stable adsorption sites for Li, Na, Mg and Ca atoms over B-graphdiyne.

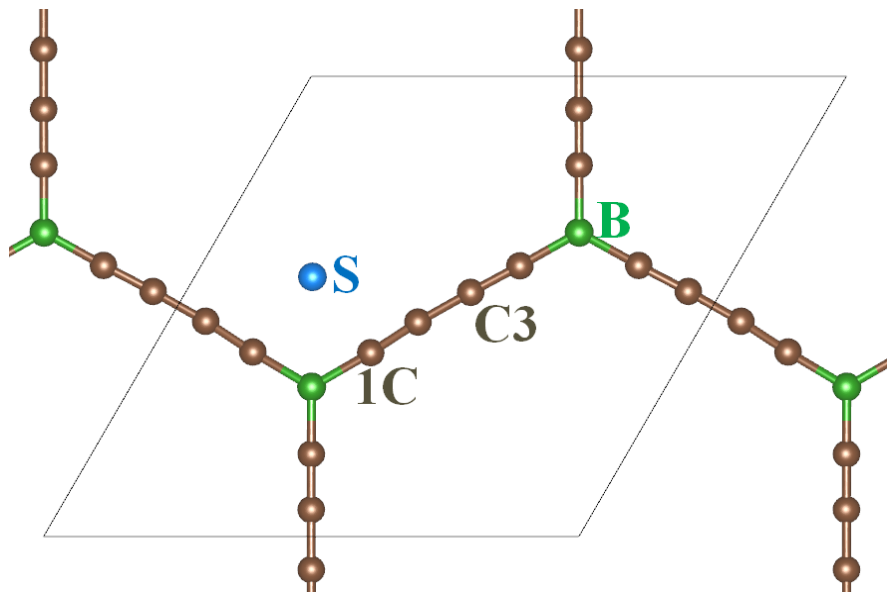

Fig. S1, Atomic structure of B-graphdiyne monolayer.

Table S1, Predicted most stable adsorption sites for the single Li, Na, Mg and Ca adatoms over the single-layer B-graphdiyne. Here, $E_{ad}$, $L_{x-y}$, Z and ΔQ depict, respectively, the corresponding adsorption energy, distance between the closest x and y atoms, the out-of-plane movement of an adatom at the "S" adsorption site (shown in Fig. S1) and the charge transfer from a single adatom to the B-graphdiyne monolayer predicted by the Bader charge analysis.

| Most stable adsorption sites | Li | Na | Mg | Ca |
|---|---|---|---|---|
| First | S site | S site | S site | S site |
|  | $E_{ad}$= -1.7 eV | $E_{ad}$= -1.75 eV | $E_{ad}$= 0.65 eV | $E_{ad}$= -1.22 eV |
|  | $L_{B-Li}$= 2.453Å, Z=0.0 Å | $L_{B-Na}$= 2.9 Å, Z=0.0 Å | $L_{B-Mg}$= 2.5 Å, Z=1.38Å | $L_{B-Ca}$= 2.627 Å, Z=1 Å |
|  | ΔQ= 0.994 \|e\| | ΔQ= 0.993 \|e\| | ΔQ= 1.467 \|e\| | ΔQ= 1.469 \|e\| |
| Second | 1C top | B top | B top | C3-C3 bridge |
|  | $E_{ad}$= -1.22 eV | $E_{ad}$= -1.34 eV | $E_{ad}$= 0.73 eV | $E_{ad}$= -0.94 eV |
|  | $L_{1C-Li}$= 2.054 Å | $L_{B-Na}$= 2.431 Å | $L_{B-Mg}$= 2.388 Å | $L_{C3-Ca}$= 2.33 Å |
|  | ΔQ= 0.988 \|e\| | ΔQ= 0.992 \|e\| | ΔQ= 1.252 \|e\| | ΔQ= 1.4 \|e\| |
| Third | B top | 1C top |  | B top |
|  | $E_{ad}$= -1.19 eV | $E_{ad}$= -1.32 eV |  | $E_{ad}$= -0.85 eV |
|  | $L_{B-Li}$= 2.106 Å | $L_{1C-Na}$= 2.45 Å |  | $L_{B-Ca}$= 2.343 Å |
|  | ΔQ= 0.989 \|e\| | ΔQ= 0.992 \|e\| |  | ΔQ= 1.44 \|e\| |